\documentclass[12pt, preprint]{aastex}
\usepackage{emulateapj5}
\newcommand\ASCA{{\it ASCA}}
\newcommand\asca{{\it ASCA}}
\newcommand\rosat{{\it ROSAT}}
\newcommand\ROSAT{{\it ROSAT}}
\newcommand\psc{\ifmmode{\rm\,cm^{-2}}\else{${\rm\,cm^{-2}}$}\fi}

\slugcomment{To appear in the Astrophysical Journal Supplement Series, April 2001}

\begin{document}
\title{The Seyfert-Starburst Connection in X-rays. I. The Data}
\author{N. A. Levenson, K. A. Weaver\altaffilmark{1}, and T. M. Heckman}
\affil{Department of Physics and Astronomy, Bloomberg Center, Johns Hopkins University, Baltimore, MD 21218}
\altaffiltext{1}{Laboratory for High Energy Astrophysics, Code 662, NASA/GSFC, Greenbelt, MD 20771}
\shorttitle{The Seyfert-Starburst Connection in X-Rays}
\shortauthors{Levenson et al.}

\begin{abstract}
We analyze X-ray spectra and images of a sample of Seyfert 2
galaxies that unambiguously contain starbursts, based on
their optical and UV characteristics.
Although all sample members contain
active galactic nuclei (AGNs), supermassive black holes or other related
processes at the galactic centers alone cannot
account for the total X-ray emission in all instances.  
Eleven of the twelve observed galaxies are significantly resolved with
the \rosat{} HRI, while six of the eight sources observed with the 
lower-resolution PSPC also appear extended on larger scales. 
The X-ray emission
is extended on physical scales of 10 kpc and greater, which we
attribute  to starburst-driven outflows and supernova-heating of the interstellar medium.
Spectrally, a physically-motivated composite model of the X-ray emission that includes 
a heavily absorbed ($N_H > 10^{23}{\rm\, cm^{-2}}$)
nuclear component (the AGN), power-law like 
scattered AGN flux, and a thermal starburst 
describes this sample well.  
Half the sample exhibit iron K$\alpha$ lines, which are typical of 
AGNs.  

\end{abstract}

\keywords{galaxies: Seyfert --- X-rays: galaxies}
\section{Introduction}

One of the most important questions about Seyfert galaxies is
the fundamental nature of their energy source.
While accretion onto supermassive black holes
generally  describes successfully the
central engines of many Seyfert galaxies
(e.g., \citealt{Miy95}, \citealt{Tan95}, and \citealt{Nan97}),
starbursts are also important, often in the same
galaxies.  The bolometric luminosity of
circumnuclear starbursts can be comparable to
active galactic nuclei (AGNs; \citealt{Gen98}). 
Other observations suggest that the AGN and starburst 
phenomena may be directly related on cosmological scales.
At low redshift ($z < 2$), the populations of quasars
and star-forming galaxies evolve similarly (\citealt{Boy98} and
references therein), and masses of black holes in galactic centers 
are strongly correlated with the velocity dispersions 
of their ``host'' stellar spheroids \citep{Geb00,Fer00}.

Accretion alone does not fully account
for the variety of emission observed in AGNs.
In particular, much of the optical and ultraviolet emission from
type 2 Seyfert galaxies is a featureless continuum
that {\it cannot} be light from an obscured Seyfert 1
nucleus that is scattered off dust or warm gas,
as optical spectropolarimetry (Tran 1995a,b,c)\nocite{Tra95a,Tra95b,Tra95c} 
and UV spectroscopy \citep{Hec95} demonstrate.
Two possibilities have been advanced for the origin of the featureless
continuum: 
optically-thin thermal emission from warm 
gas in the reflection region (Tran 1995a,b,c)\nocite{Tra95a,Tra95b,Tra95c}  
or light from a circumnuclear starburst
\citep{Cid95,Hec95}.
UV images and high-quality optical and UV spectra of a flux-limited sample of
type 2 Seyfert nuclei \citep*{Hec95,Hec97,Gon98,Gon00}
strongly suggest that there are
two types of Seyfert 2s. In roughly half
the sample nuclei,  a strong UV continuum shows the
unambiguous spectroscopic signature of hot stars.  
These are the Seyfert/starburst (Sy2/SB) composite galaxies.
In the other members of the sample,
the UV continuum is relatively faint and sometimes has a cone-like
structure aligned with the radio axis. In these latter, the nuclear
UV light is probably due to a combination of scattered light and
optically-thin thermal emission from warm ($10^5$ to $10^6$ K) gas in the 
reflection region.

X-ray spectroscopy is a valuable tool for investigating
the Seyfert-starburst connection.
The soft X-ray emission in about one-third of Seyfert 2 galaxies
is thermal \citep{Tur97}, and this emission is 
consistent with the soft X-ray properties 
of nearby starbursts 
\citep*{Dah98,Wea00}.
At energies of 1--2 keV, 
emission from X-ray binaries and individual supernova remnants
dominates in starbursts, while at lower energies,
thermal emission from diffuse gas that has been heated
to temperatures of 0.2 to 1.0 keV by the collective
effect of stellar winds and supernovae 
\citep[e.g.,][]{McK77}
is the primary source of X-ray emission.
This hot gas can escape the starburst in the form of a galactic
``superwind'' \citep{Che85,Hec93}.

Because hard X-rays can penetrate large column densities of
gas and dust, they serve as direct observational probes of
deeply buried Seyfert 1 nuclei in Seyfert 2 galaxies, which
the unified model suggests. (See \citealt{Ant93} and \citealt{Urr95}
for reviews.)
In the unified model, the X-ray emission 
in a type 2 Seyfert nucleus consists of the
sum of the heavily-absorbed power-law emission from the
hidden Seyfert 1 that has been transmitted through the
torus and the lightly-absorbed power-law emission that has been
reflected off the mirror.  
In the composite objects, both the buried Sy 1 and the starburst produce
X-rays.
While the AGN dominates the Sy2/SB spectra at high energies, 
below around 2 keV the reflected and transmitted
X-ray emission of the AGN
are combined with the soft thermal spectrum
of the starburst component.

This is the first of two papers in which we investigate the X-ray
signatures of the Seyfert-starburst connection.  In this work, we
discuss our analysis of the X-ray images and spectra of a sample of
known Sy2/SB composites.   We
present X-ray images and spectra obtained with both \rosat{} and
\asca{} of this sample in \S\ref{sec:data}.  We do not expect the
unresolved active nucleus to be the sole source of X-ray emission in
each galaxy and examine the images for extended emission
(\S\ref{sec:image}).  We fit the spectra with complex models that are
physically motivated (\S\ref{sec:specfit}).  We comment on individual
galaxies in \S\ref{sec:indiv} and draw conclusions in \S
\ref{sec:concl}.
In Levenson, Weaver, \& Heckman (2000; Paper
II)\nocite{LWH00j}, we compare the X-ray properties of these
composites to ``pure'' starbursts and Seyferts and discuss the
implications for the connection between AGNs and starbursts.

\section{The Data\label{sec:data}}
\subsection{Selection Criteria\label{subsec:select}}
For our study, we have chosen galaxies from the
\citet{Hec95} sample.
In order to investigate how the starburst phenomenon relates to Seyfert 2s,
these authors identified the 30 brightest Seyfert 2 nuclei
based on [O {\sc iii}] $\lambda\lambda 4959+5007$ emission line flux
and nuclear nonthermal monochromatic flux ($\nu F_\nu$) 
at 1.4 GHz from the compilation of \citet{Whi92}.
Members of the \citet{Hec95} sample have either 
$\log F_{\rm [O III]} \ge -12.0 {\rm\, erg\,cm^{-2}\,s^{-1}}$,
or $\log F_{1.4} \ge -15.0 {\rm\, erg\,cm^{-2}\,s^{-1}}$, or both.

A starburst is evident in approximately half of these bright Seyfert 2 
galaxies.  They have 
the spectroscopic signatures of luminous young stars, such as
stellar wind lines, Balmer absorption,
and broad Wolf-Rayet emission features 
(\citealt*{Hec97,Wan97,Cid98}; \citealt{Gon98,Sch99,Gon00}). 
We restrict our sample to these known Seyfert/starburst composites
plus several
galaxies that \citet{Hec95} 
had excluded from their study only because IUE had not
observed them but which fulfill the original selection criteria.
Our sample is unbiased, but we exclude several known composite
galaxies that were not part of the original \citet{Whi92}
sample or that fail to meet the brightness criterion.

The optical 
properties of the sample are listed
in Table \ref{tab:info}.  The objects are listed in order of
right ascension (J2000).  Distances (column 4) are calculated assuming
$H_0 = 75 {\rm \, km\,s^{-1}\,Mpc^{-1}}$.  The scale (in pc) is equivalent
to $1\arcsec$ (column 5).  
The galaxy inclinations (column 7) should be interpreted cautiously.
They have been determined assuming circular disk geometry, yet many of
these are disturbed systems, for which this method is inappropriate.
The references for Galactic column density, inclination, and starburst
activity for each galaxy are noted.
Table \ref{tab:obs} contains information 
on the X-ray observations we present here, which we describe in detail
subsequently. 

\subsection{\rosat{} HRI Observations}
The \rosat{} High Resolution Imager (HRI) is a particularly 
appropriate tool for identifying extended nuclear emission.
In addition to the advantage of its absolute spatial 
resolution, its
sensitivity is greatest at the soft energies typical of the 
widespread thermal emission of starburst galaxies.
We have obtained HRI data for all 14 members of
the sample, although two of the exposure times 
were too short to provide any useful information (Mrk 463 and NGC 7674).
We use an off-source region in each field to determine the
constant background to subtract.  The background-subtracted HRI count
rates are listed in column 4 of Table \ref{tab:obs}.  

Figure \ref{fig:hriall} illustrates the 
relationship of the soft X-ray
and optical emission in these objects and the radial distribution of the
soft X-rays.  The HRI images are
constructed from the raw ($0\farcs5$) pixels, smoothed with
a two-dimensional Gaussian of FWHM$=4\arcsec$.  The alignment
is based on the \rosat{} pointing and astrometry of the 
Digitized Sky Survey (DSS) frames.  When additional point
sources are located within the complete observed field,
they are used to improve the alignment. 
Because members
of the sample themselves may be extended X-ray sources, we do not
use them to correct the astrometry.
The optical and X-ray alignment of Mrk 1073 
is uncertain.  Based on the \ROSAT{} telescope pointing alone, the X-ray source
appears offset 25\arcsec\ from the optical location of Mrk 1073.
No other sources are detected in the field, so we have 
approximated the alignment of the images.

We extracted images at the full spatial resolution of the HRI in
order to identify extended emission, 
comparing the
observed radial profile with the theoretical point spread function 
(PSF; \citealt{Dav99}).
A two-dimensional Gaussian fit to each smoothed image
defines the center of emission.
We combined the raw $0\farcs 5$ pixels into $3\arcsec$ bins to 
construct the radial profile,
fitting the amplitude and background to the data.
In no case is the target galaxy or any other source
in the field bright enough to correct for errors in the telescope
pointing \citep[cf.][]{Mor94}.  
When known point sources appeared in
the field, however, we compared their observed radial profiles to
check the consistency of our results.

\subsection{{\it ROSAT} PSPC Observations}
\rosat{} PSPC observations provide spatial 
and  spectral information on some of these Sy2/SB composites.
All the sources were the specified central targets
of the observations, so they are located at the centers of
the observed PSPC fields and do not require vignetting corrections.
We subtracted a constant background, measured over a large central region excluding the 
source, to determine the count rates listed in column 7 of Table \ref{tab:obs}.

PSPC contours overlaid on optical images are shown for Mrk 463 and NGC 7674,
the sources for which we lack adequate HRI data 
(Figures \ref{fig:poptm463} and \ref{fig:poptn7674}).
Figure \ref{fig:pspcall} contains the 
PSPC radial profiles.  
We use the same technique described for the HRI data above
to identify extended soft X-ray emission in the PSPC observations,
applying the appropriate theoretical point spread function, assuming
typical energy of 1 keV 
\citep{Has92}.  

We examined the soft
(0.2--0.4 keV) and hard (0.4--2.0 keV) PSPC bands separately, as well as the
total emission.  We present these broad-band images  
(Figures \ref{fig:phsn1068}--\ref{fig:phsn7582}) in the cases in 
which we measure significant differences.

Spectra were extracted from circular regions of 
$2\arcmin$ radius typically, except in crowded fields.  
Some sample members contain double nuclei, which
are separated by a few arcseconds
or less and are therefore unresolved.
Thus, the contributions of both nuclei are combined in 
the resulting spectra.
Nearby,
source-free regions of the sky were used to determine 
the background.
The data are grouped so that each bin has a minimum of 30 counts,
and therefore $\chi^2$ statistics are appropriate in the model fits.
Spectral models were fit from 
0.2 to 2.0 keV
using the appropriate response matrices.  The 
low-energy sensitivity of the PSPC
is particularly useful to detect the soft
thermal emission of the starburst and to constrain the 
absorbing column density along
the line of sight.

\subsection{\ASCA{} Observations\label{sec:dataasca}}
We use the \asca{} observations of these composite galaxies to
identify their AGN and starburst components spectrally. 
See \citet{Tan94} for a description
of the \ASCA{} experiment.
We processed the \ASCA{}  data to exclude times of 
high background rates.  The accepted data
were taken outside of and greater than 16 s 
after passages through the South Atlantic Anomaly,
at elevations above the Earth's limb greater than $5^\circ$, 
at times greater than 16 s after a satellite day/night transition, 
and with geomagnetic cutoff rigidity greater than $6 {\rm \,Gev\,c^{-1}}$.
Source counts were extracted from circular regions having a radius 
approximately 
$3\farcm2$ on the SIS detectors (SIS0 and SIS1) 
and $4\farcm4$ on the GIS detectors (GIS2 and GIS3),
appropriate to the instrumental resolution.
Source-free regions of each detector were used for the corresponding
background measurements.  The \asca{} count rates and SIS mode
are listed in columns 10 and 11 of Table \ref{tab:obs}.
The \asca{} spectra were binned according to the same criterion as the 
PSPC spectra.
We include SIS data for energies in the range 0.6 to 10.0 keV, and
GIS data in the range 0.7 to 10.0 keV.
We model the spectra with the latest \asca{} effective
area calibration.  
All models include a constant factor to account for the 
few percent known difference in the SIS and GIS fluxes.  The fluxes
and normalizations listed in the tables are for the SIS0 detector.

Several of the galaxies are variable (NGC 5135, Mrk 463, NGC 6221, and NGC 7582),
typically showing flux changing by factors of less than 50\% 
on timescales on the order of $10^4$ s.
Our concern here, however, is using high signal-to-noise data to determine
the net properties of these galaxies, so we
use the time-averaged spectra for each pointing. 
Furthermore, most of the variability is evident at high energies
and does not affect the soft starburst emission that we emphasize. 
Mrk 477 was observed on two separate occasions.  We found no
variation between these observations, so we have combined
the two data sets.  NGC 7582 was also observed twice.  The
spectrum varied significantly around energies 
$2\lesssim E \lesssim 3$~keV \citep{Xue98}, so we have
treated the two data sets independently.  The soft X-ray
emission in the \asca{} bandpass does not vary, so we
have used the same PSPC spectrum in the joint fits of data
from both satellites.

\section{Imaging Results\label{sec:image}}

\subsection{\rosat{} Images\label{subsec:rosimg}}
We summarize the results of detecting extended X-ray emission 
in the \rosat{} data in Table \ref{tab:extend}.
We use the maximum radius at which residual flux is detected
to approximate the physical scale of the extended emission
in HRI and PSPC observations (columns 2 and 5, respectively).
Subtracting the best-fitting theoretical PSF from the observed
values, we estimate the fraction of total flux that is in excess of a
single, central point source using two techniques.
The radial profile plots (Figures \ref{fig:hriall} and \ref{fig:pspcall})
and columns 3 and 6 of Table \ref{tab:extend} 
contain the results of the first method,  
in which we fit the entire PSF weighted by
the Poisson errors.  These residuals tend to be extremely
small or negative at the smallest radii and increase at larger
radii. The residual profile corresponds to the intrinsic shape of the 
non-AGN flux, which here implies rings of emission in the two-dimensional
images. 
Such unfilled rings are unlikely to generally represent the true physical
structure of the extended components.  Thus, this method 
overestimates the unresolved emission and yields
conservative estimates of the minimum extended fraction in each galaxy.
The associated errors printed in the table are derived from 
Monte Carlo simulations of this
technique and do not reflect its systematic bias in favor of
the AGN contribution.
In the second method (columns 4 and 7 of Table \ref{tab:extend}), 
we require that the
residual distribution be flat within the central core of 
the PSF.  This constraint is more realistic when
truly extended emission is present and also fits unresolved sources well.
A disadvantage is that the errors associated with this technique are 
somewhat larger than those of the first method. 
As a result, the HRI observation of NGC 7130, for example, is resolved 
according to the first estimate, but we
determine only an upper limit on the extended fraction using the more realistic
model. 

The X-ray emission from three galaxies (Mrk 1066, Mrk 273, and Mrk 477) 
is unresolved in the HRI data, according to the conservative estimates.
Using the constrained model, Mrk 273 and Mrk 477 are resolved.   
Only in the case of Mrk 1066 do we fail to find any evidence for extended emission.
In all other galaxies observed with the HRI, the soft X-ray emission cannot be attributed to
a single point source alone.
At the lower spatial resolution of the PSPC, six galaxies are
extended: NGC 1068, Mrk 78, NGC 5135, Mrk 266, NGC 7582, and NGC 7674.
The counting statistics as well as the intrinsic resolution are particular
limitations in the observations of Mrk 78 and NGC 7674.  Mrk 78 is significantly
resolved only according to the constrained (second) method. 
In the case of NGC 7674, only using the first method are the errors
small enough to determine that the emission is extended with confidence.

The HRI and PSPC are sensitive to extended emission on different angular scales.
The resolution of the PSPC is around 30\arcsec, so it is suited for 
measuring large angular scales.
The resolution of the HRI is around 5\arcsec, but it is not as sensitive as the PSPC.  It
cannot detect the extremely low surface brightness of the most extended emission,
such as that due to a starburst-driven superwind, but it does resolve brighter,
more distant sources.  While our results suggest that
the PSPC is optimized
for detecting extended emission at radii of 50\arcsec, the HRI generally only detects
extended emission out to radii of 20--30\arcsec.  Therefore, the combination of
resolution, sensitivity and varying galaxy distances biases the physical scales
on which we can actually measure extended emission with the PSPC and HRI.

The galaxies that are significantly resolved in the total PSPC
images also exhibit distinct morphology in the soft and hard energy
regimes. 
The softest emission from NGC 1068 is 
not very centrally concentrated 
(Figure \ref{fig:phsn1068}a).  The emission profile is flat,
not sharply peaked.  At harder energies,
NGC 1068 is strongly centrally concentrated, and the low surface brightness
emission extends toward the northeast and southwest (Figure \ref{fig:phsn1068}b).
Although the PSF of the PSPC has a broader core
at lower energies, this energy dependence alone does not account for
the observed difference.  We detect truly distinct morphologies 
in the soft and hard emission. 
The hard emission from NGC 5135 is symmetric and
centrally concentrated, while in this galaxy the soft emission is
more diffuse and irregular (Figure \ref{fig:phsn5135}).
The low-energy image of Mrk 266 consists of two distinct sources (Figure \ref{fig:phsm266}a).  
The galaxy's nuclei account for the southern component, while the 
additional extended emission also detected in the HRI observation 
coincides with the northern source.  Only the southern component is significant
at higher PSPC energies (Figure \ref{fig:phsm266}b).  The southern component 
is resolved in both energy bands, supporting its identification as the compound
nucleus.
The soft emission of NGC 7582 (Figure \ref{fig:phsn7582}a) is extremely diffuse
and extends mostly perpendicular to the plane of the galaxy, as expected from
a superwind.  
In this case, the hard emission is sharply peaked at the nucleus and extends primarily
in the galactic disk (Figure \ref{fig:phsn7582}b).

\subsection{{\it ASCA} Images\label{subsec:ascaimage}}
We constructed images from the \asca{} data in two noteworthy cases: 
Mrk 273 and NGC 7582.  Both have X-ray ``companions'' that appear
nearby projected onto the plane of the sky but are
not physically associated with the galaxies of interest.
Mrk 273 and its companion, Mrk 273X, separated by about $2\arcmin$, are clearly 
resolved by the HRI and PSPC (Figure \ref{fig:m273x4}).  
The latter contributed about 
one-third of the soft X-ray flux in the PSPC observation of
1992 May \citep{Tur93} and HRI observation in 1992 June, but 
faded significantly by 1994 December, when \asca{} observed the field.
We can distinguish Mrk 273 and Mrk 273X in the SIS data because the
point spread function has a sharp central core.  The
companion contributes less than 10\% of the total SIS0 counts, mostly
in the soft (0.5--2 keV) band.
NGC 7582 and its companion are also separated by about $2\arcmin$
and resolved in HRI \citep{Sch98} and PSPC images (Figure \ref{fig:n7582x4}).
This companion is much fainter than NGC 7582, 
contributing only 15\% of the total counts.  It is relatively stronger
at soft energies, having 30\% of the flux of NGC 7582 in the SIS0 detector.
Note that in both of these cases, the companions do not significantly
affect the results of spectral modelling.

\section{Spectral Modelling\label{sec:specfit}}
While fitting spectral models to the data, we are ultimately guided by
the realistic physical conditions in the sample galaxies.  While the
AGNs emit characteristically strong X-rays, we know that these are
not the only X-ray sources present.  The images clearly reveal extended
soft X-ray emission, which cannot be due to the active nuclei.  
Because all the sample members definitely contain
starbursts, we expect thermal X-ray emission from these,
by analogy with ordinary starburst galaxies that lack AGN.

We develop the complex models that account for the multiple
physical sources gradually, however, beginning with several simple models that we
apply to all the spectra.
Although not all models are realistic for every galaxy, by systematically 
building complex models out of the likely components, we demonstrate
which pieces are significant in different cases.  Also, the
uniform set of models facilitates comparison of the sources,
even in the instances where we obviously require multi-component models
to achieve good fits, such as NGC 1068.
We present model fits to \asca{} data first and then joint fits with the PSPC
data, when available.
For Mrk 78 and NGC 7674, which lack \ASCA{} data, only PSPC fits 
are presented.

\subsection{Power-law Models}
We first examine a single absorbed power law, where the photon 
index and absorbing column density are left as free parameters, 
although the absorption is constrained to be greater than or 
equal to the Galactic value.  
This sort of model tends to  
approximate well the spectra of Seyfert 1 galaxies. For 
most of the composite objects,  however, this model (Model 1) either fits the
data poorly or produces unphysical results (Table \ref{tab:pl}).
The derived photon indices tend to be small and the model generally 
fails to account for the observed soft X-ray emission.  None of
the single power law fits to \asca{} spectra are acceptable.

To better model the soft emission, we try various solutions that are 
physically motivated.  First, we examine adding a second power 
law component (Model 2), which represents the case of a buried Seyfert 
nucleus with material that scatters a significant amount of the 
blocked soft X-ray emission in our direction:  
\begin{eqnarray}
I({\rm photons\ s^{-1} \,cm^{-^2}\,keV^{-1}}) = 
\exp(-N_{H0}\sigma_{abs})\nonumber\\
\left(\exp(-N_{H1}\sigma_{abs})(A_1E^{-\Gamma})+ A_2E^{-\Gamma}\right),
\end{eqnarray}
where $N_{H0}$ and $N_{H1}$ are the Galactic and nuclear column densities,
respectively, $\sigma_{abs}$ is the absorption cross-section,
$A_1$ and $A_2$ are the magnitudes of the intrinsic and scattered
components, respectively, and $\Gamma$ is the photon index.
In this case, the
soft X-ray photon index is set equal to the hard X-ray (nuclear)
index, the column density toward the scattered component is fixed 
at the Galactic value, and the nuclear column density is a free  
parameter.  In addition, we fix the photon indices at 1.9, the 
mean value for the {\it intrinsic} index for Seyfert 1 galaxies
\citep{Nan94}.
With the exception of NGC 1068 and NGC 6221, all objects prefer 
this model over the single power law, 
with a distinct, heavily absorbed component 
having $N_H > 10^{23} {\rm\,cm^{-2}}$ (where $N_H = N_{H0}+N_{H1}$)
indicated in all cases, either by direct observation or inferred by 
the presence of strong Fe K$\alpha$ fluorescence, which is
produced in a region of high column density.
In addition, 
the extra soft X-ray power law component accounts for most of the  
residual error.  Several fits are very good; for example, 
$\chi^2/{\rm dof} \approx 1.1$ for Mrk 1066, Mrk 273, and Mrk 463
(Table \ref{tab:pl2}).

\subsection{Thermal Plasma Models}
Next we examine a model that replaces one of the power law
components with a thermal plasma having solar abundances (Model 3). 
Physically, this represents the soft X-ray emission due to starburst
activity.  In addition to the possible transmitted and reflected AGN components,
some source of emission that can be extended is required for consistency 
with the soft X-ray emission resolved with \rosat.
For the thermal component we use the so-called ``MEKAL'' model of
Mewe and Kaastra \citep{Mew85,Arn85,Mew86,Kaa92}, with updated Fe L
calculations \citep{Lie92}.  Our intention is to determine
whether a thermal model better describes the soft X-ray emission
than a power law model.   Model 3 yields a better fit than Model 2
for 6 of 13 ASCA observations and for 4 of 
7 cases where PSPC data are available, for a total of 7 galaxies
(Table \ref{tab:plm}).
For cases where the fits are worse, they can
be improved by reducing the metal abundance, but since abundances
are difficult to estimate from data of this quality, we do
not pursue this option.

Finally, we examine three-component models that 
combine Model 2 with thermal emission.
In addition to the original power law, this model (Model 4) contains
a second power law,
to represent scattered flux, and a thermal 
component to model the soft X-ray emission, which might be
expected from a combination of an AGN and starburst.
Table \ref{tab:pl2m} lists these results only
for the
cases where this model significantly improves the fit.
The three-component model improves the fits 
over Models 1 or 2 significantly 
in 5 out of 13 cases for \asca{} fits and 
in 6 out of 7 cases for joint fits, for a total of 6 out of
12 galaxies with \asca{} data.  

\subsection{Best-Fitting Models}
Table \ref{tab:best} contains a summary of the best-fitting model for each galaxy,
including a Gaussian to model the Fe K$\alpha$ line (discussed below) 
in cases where it is detected.  
For non-detections, upper limits on the line equivalent widths (EWs) are noted.
In each case, the inclusion of additional components or variable parameters
over models with fewer free parameters 
is statistically significant at the 90\% confidence level, based on an 
$F$ test.
The data and model fits are displayed in Figure \ref{fig:spectra}.
Although we used both SIS and both GIS detectors independently in the
model fitting, we plot the average of data from the SIS detectors
and the average of data from the GIS detectors separately for clarity.
Both \rosat{} and \asca{} spectra are shown where they are available.

Although the spectra of Mrk 1073 and IC 3639 are of low statistical 
quality, we can use their images to infer which process dominates the soft 
X-ray spectrum.  Throughout the sample, the objects with the strongest thermal 
spectra, such as NGC 5135, tend to be most extended.  The soft X-rays are resolved in 
Mrk 1073 and IC 3639, so the most likely spectral model for both is 
thermal emission.  Specifically,
because Mrk 1073 is not a point source, we have adopted Model 3 as the best fit,
but Model 2 also formally fits well and explicitly illustrates the heavily
obscured ($N_H\sim 10^{24}\psc$) AGN.
In the case of IC 3639, although statistical improvement of the 
complex model (Model 4) over the preferred Model 3 is not significant, 
we learn from
it that the scattered contribution in this galaxy is relatively
weak.  Hence, the single power law we observe in this best-fitting model 
is the intrinsic nucleus.

We experimented with several modifications to the basic models.
For example, NGC 1068 has an extremely complex spectrum 
\citep*{Uen94,Iwa97},
and we require several
emission lines and multiple thermal components to fit it adequately.
The PSPC is very sensitive to small changes in $N_H$, 
and increasing the low-energy column density improves some  
of the fits.  This is the case for NGC 7582, so we have allowed 
the soft $N_H$ to be free for all fits to this galaxy.  Mrk 273 also prefers 
absorption that is slightly higher than the Galactic value, but since 
this does not affect our results in a significant manner, we
use only the Galactic value for $N_H$ in the fits presented here.

We have only PSPC spectra of Mrk 78 and NGC 7674.  We considered
several models of their soft emission, including a power law, thermal
emission, and combinations of the two.  In each case, a single
component with Galactic absorption best fits the data.  
This component is a thermal plasma in Mrk 78 and a power law in NGC 7674.
The parameters of these models are listed in Table \ref{tab:pspc}.

\subsection{Fe K$\alpha$ Emission}

Fe K$\alpha$ emission is common to active galaxies.
In many cases, particularly Seyfert 1s, the lines are broadened by Doppler
and relativistic effects, indicating that a significant component
of the line originates in the inner
regions of an accretion disk \citep{Nan97}.
Seyfert 2 galaxies, on the other hand,
often show narrow features, which appear to originate farther out in the 
galaxy, perhaps due to fluorescence of the nuclear continuum in the 
obscuring torus \citep{Wea96}.  In any case, the profile and
variability of the iron line can tell us a great deal about the physics 
of the gas that is fueling the AGN.

Several members of our Seyfert/starburst sample exhibit Fe K$\alpha$ lines.
Since the iron lines are not the focus of this paper, we consider modeling 
the lines only in the best-fitting of the previous models  
(Table \ref{tab:best}).
Typically, the Fe K lines are narrow and can be fitted with an unresolved 
Gaussian with $\sigma < 0.05$ keV.  
We measure equivalent widths 
with respect to the power law 
component responsible for the hard X-ray emission.
For cases where this component physically represents 
the intrinsic (yet absorbed) power law, the EW ranges from approximately
100 eV to 1 keV.  
In several cases (e.g., NGC 1068, Mrk 1066), however, this 
intrinsic AGN is not detected directly because it is completely absorbed.  
In these galaxies, the only detectable hard X-ray continuum is
due to the scattered power law.
As a result, measuring these EWs with respect to 
the genuinely weaker scattered (rather than intrinsic) AGN,
we find these EWs to be large, greater than 3 keV. 
In these cases, we crudely estimate the line EW with respect to the 
intrinsic AGN continuum, assuming the mean fraction of intrinsic emission
that is measured to be scattered in the other sample members indicates
the true intrinsic component.
This average scattering fraction $f_{scatt}=0.05$.
Thus, increasing the
continuum by a factor of 20, we predict the EW with respect to
the intrinsic continuum to be 50 and 160 eV in NGC 1068 and Mrk 1066,
respectively.  The corresponding upper limits are 
EW $< 50$, 470, and 270 eV in IC 3639, Mrk 266, and NGC 7130, respectively.

NGC 6221 is unusual in that it contains a broad line with EW$\approx 400$ eV.  
We find $\sigma = 0.5 (+0.4, -0.02)$ keV with 90\% confidence.
The Fe line properties of this galaxy are more characteristic of Seyfert 1s,
where the reprocessing material is close to the galaxy nucleus.
The absorption of the hard component in NGC 6221 is only 
$10^{22}{\rm\,cm^{-2}}$, which is much lower than in any other
members of the sample.  As a result, 
although this galaxy possesses  both the scattered power law and
thermal soft X-ray emission typical of other Sy2/SB composites, 
the intrinsic AGN dominates the observed X-ray emission at low 
and high energies during the \asca{} observation.

The strength of the Fe K line is related to the 
column density of the fluorescing material in a predictable fashion. 
The uncertainty in studying this relation has historically been due a
lack of reliable measurements of $N_H$, especially for very large column 
densities ($N_H \gtrsim 10^{23} \psc$).  
In Figure \ref{fig:ewnh}, we plot Fe K EW versus
$N_H$ for our sample (solid symbols).  By measuring 
the line with respect to the 
intrinsic continuum, where we account for both the absorbed  
X-ray emission from the AGN and the scattered flux,
we find that the EWs are similar to those expected for 
the inferred column densities.  For comparison, we also plot the results from
Turner et al. (1997\nocite{Tur97}; open symbols), 
which were derived using the observed 
continuum.  Theoretical predictions for spherically-symmetric and
torus geometries are  plotted with the dashed and
dot-dashed lines, respectively.  The \asca{} data are of too poor quality to be 
able to distinguish between these models, but we conclude that our 
measurements are consistent with such geometries and that these objects indeed
possess buried Seyfert 1 nuclei.

\section{Notes on Individual Galaxies\label{sec:indiv}}
\subsection{NGC 1068}

NGC 1068 is a member of a group \citep{Gar93} and contains a stellar bar
\citep{Thr89}. 
Broad optical emission lines are observed in polarized light \citep{Ant85}.
The X-ray emission from NGC 1068 has been studied in detail 
\citep{Wil92,Uen94,Iwa97,Net97}, and its spectrum is extremely complex.  

For consistency, we present the simple model fits to the
spectrum, although none of these are satisfactory.
The best-fit model includes a power law, two thermal components,
and three Gaussian components.  The intrinsic power law of the
AGN is completely absorbed.  In these joint fits of \asca{} and
PSPC data, we have rejected SIS data at energies $E<0.8$ keV, because
the SIS calibration of such bright, soft sources at low energies
is highly uncertain \citep{Wea00b}.
We find that 80\% of the
soft X-ray emission is thermal. 

NGC 1068 is clearly extended in X-ray emission, and the 
soft and hard PSPC bands are distinct.
Although a radio jet has been observed \citep{Wil82}, 
the extended X-ray emission is associated with the starburst,
not the jet \citep{Wil92}.  Measured both with the HRI and PSPC,
the soft X-ray emission extends on scales greater than 7.4 kpc.

\subsection{Mrk 1066}
Mrk 1066 has a bar but is not a member of a group or cluster. 

In the spectrum of Mrk 1066, we observe a single power law component, 
soft thermal emission, and an Fe line.  
The thermal emission accounts for half
the total soft X-ray flux. 
The absorption of the power law is consistent with the Galactic
column density; this component is due to the scattered AGN.
We do not detect the intrinsic hard AGN emission directly, 
even at the highest energies of the \asca{} bandpass.
The AGN is heavily absorbed, with $N_H > 10^{24}\psc$.  
Because the equivalent width of the line is measured with respect
to the scattered component, it is large: EW $\approx 3000$ eV.
\subsection{Mrk 1073}
Mrk 1073 is part of the Perseus cluster and paired with UGC 2612.
It contains a central bar.

This galaxy lies at the edge of the cluster, complicating
background subtraction of the X-ray spectra.  We have selected
background regions preferentially away from the cluster emission.
The signal-to-noise ratio of the SIS1 data is very low, so
we have not used these data in the spectral fitting.
The best-fitting spectrum consists of a single power law and a thermal
contribution.  Most (91\%) of the soft X-rays are thermal. 
Although a second power law component instead of 
thermal emission fits the soft X-ray data equally well, we 
prefer the thermal model because Mrk 1073 is an extended X-ray
source, as observed with the \ROSAT{} HRI. 
The physical scale of this extent is 9.5 kpc. 

\subsection{Mrk 78}
Mrk 78 is a barred galaxy.  Its elongated radio and
[\ion{O}{3}] emission are correlated \citep{Ped89}.

The PSPC spectrum of Mrk 78 is best fit with a thermal plasma,
$kT = 0.76$ keV, and Galactic absorption. 
Observed with the HRI, the soft X-ray emission is extended,
to a radius of 17 kpc.  The PSPC data also suggest extended emission,
on a scale of 36 kpc.
\subsection{IC 3639}
IC 3639 has a bar \citep{Mul97}
and is paired with ESO 381-G009.

We excluded the low-quality SIS1 data from the spectral fitting.
The remaining data are best fit with 
a two-component model of a power law and thermal emission.  
The thermal temperature is high, $kT = 2.3$ keV.
In this case, all the soft X-ray emission is thermal. 
The three-component model, including the scattered power law,
also fits the data.  In this case $kT = 1.7$ keV, but the
improvement to the fit with inclusion of additional parameters 
is not statistically significant. 
The X-ray emission is extended in the HRI data, with a radius of 6.8 kpc.
\subsection{NGC 5135}
NGC 5135 contains a bar, 
which has also been observed in the
near-infrared \citep{Mul97}, and it is in a group \citep{Kol89}.

NGC 5135 is best fit with the three-component model of a scattered
power law and thermal emission.  The soft X-ray flux from the
thermal component is slightly stronger than that of the scattered power law;
54\% of the soft X-ray emission is thermal. 
The residuals of the fit suggest that an additional low-energy
($E< 0.5$ keV) source is present.  Analogous to nearby starbursts,
this is likely an additional thermal component, but we cannot measure
it significantly in the present data. 
The X-ray emission from NGC 5135 is extended, 
both in HRI and PSPC observations, on scales of 5.3 and 13 kpc, respectively,
and the soft and hard emission within the PSPC bandpass are spatially distinct.
\subsection{Mrk 266}
Optically, Mrk 266 contains two nuclei. 
Both nuclei and more extended emission 
have been observed in X-rays \citep{Wan97,Kol98}.  

In the spectral modelling, we have excluded SIS1 data.
Few counts were obtained in this detector, and the low-energy
response is poorly calibrated for the time of the
observation in 1999 May.  The GIS2 and GIS3 data were combined
to increase the signal-to-noise ratio.
The best model consists of a power law and thermal emission.
The latter accounts for 43\% of the soft X-rays. 
The Galactic column density alone accounts for the absorption of
the power law, indicating that we observe only the reflected
nucleus and do not detect the buried (under $N_H>10^{24}\psc$) AGN
directly.  The upper limit on the Fe K$\alpha$ line equivalent
width is correspondingly large (EW $< 9400$ eV), 
because it is measured with respect
to the reflected rather than intrinsic source.

Mrk 266 is clearly extended in the HRI data, where we
detect the two nuclei and further extended emission out to a
radius of 22 kpc.
This galaxy is also extended at the resolution of the PSPC, in
which case $R_{kpc} = 32$, with distinct spatial structure in
the hard and soft bands.

\subsection{Mrk 273}
Mrk 273 is an ultraluminous infrared  galaxy and contains two nuclei. 

We exclude the poor-quality GIS2 data from the spectral fitting of Mrk 273.
The X-ray spectrum is best fit with a four-component model: direct and scattered 
power law from the AGN, thermal emission from the starburst,
and a line.  The thermal component provides 43\% of the soft X-ray flux. 
Including the PSPC data, we have also varied the absorbing
column of the soft component.  We significantly improve the fit by
increasing the absorption on the soft components to 
$3.4 (+2.6, -1.2) \times 10^{20}\psc$ from the Galactic value
of $9.7\times10^{19}\psc$.  The temperature of the thermal
emission and the relative contributions of the two soft
components do not change appreciably in this case.

The HRI data suggest that the X-ray emission from Mrk 273 is extended,
not confined to a point source, with $R_{kpc}= 18$.
\subsection{Mrk 463}
This galaxy has two nuclei, and one of them, Mrk 463E, has
been observed to have broad optical lines in polarized light
\citep{Mil90}. 

The three-component model of a scattered power law and thermal emission
fits Mrk 463.  Thermal emission accounts for 31\% of the soft X-rays. 
The theoretical point spread function of the PSPC does not fit Mrk 463 well,
suggesting that the X-ray emission is extended on scales of 48 kpc, but
the observed radial profile is also consistent with a point source.
\subsection{Mrk 477}
Mrk 477 has a companion 50\arcsec\ away, which is connected by
tidal tails \citep{Zwi71}. 
Broad optical emission lines are observed in polarized light \citep{Tra92}.

Mrk 477 is the only member of the sample that has been observed with
\asca{} whose spectrum is not fit best with a model that includes
thermal emission.  Rather, a scattered power law and narrow
line comprise the best-fitting model.  
When a
thermal component is added to this model, the parameters are
reasonable---$kT =0.9$ keV and $A_3 = 9.3\times10^{-6}$---but
this addition does not statistically improve the fit.
The Galactic column density toward Mrk 477 is low, similar to Mrk 273 
and NGC 7582.  In those cases, when the column density
absorbing the soft X-ray emission is a free parameter, which the
PSPC data constrain well, the thermal emission is significant.
Observed with the HRI, Mrk 477 is extended to a radius of 15 kpc, measured
with the constrained technique (\S\ref{subsec:rosimg}).
\subsection{NGC 6221}
NGC 6221 contains a bar, is in a group, and
its peculiar morphology may be due to interaction 
with NGC 6215 \citep{Pen84}.
The Sy2 identification of NGC 6221 is based on the [\ion{O}{3}]/H$\beta$ 
intensity ratio in a blueshifted component of the lines \citep{Ver81,Pen84},
although using the total optical spectrum of the nucleus, it is classified
as a starburst \citep{Cid98}.

We detect both NGC 6221 and ESO 138-1 in the \ROSAT{} HRI and \asca{} GIS
images, so the X-ray identification of NGC 6221 is secure,
in contrast to the suggestion of 
\citet{Sch95} that these galaxies are
confused for one another.

Of the Sy2/SB composites we examine here, NGC 6221 appears most like 
a Sy1 in X-rays.  The absorption is low, $N_H = 1\times 10^{22}\psc$, so
the observed soft X-ray emission includes a dominant (80\%)
contribution from the intrinsic AGN, as well as the scattered
and thermal components.  Only 5\% of the soft X-ray flux is thermal.  
The observed Fe line is broad, with
EW$= 400$ eV, centered at 6.6 keV.  The EW is relatively
large compared with  Sy 1 emission lines, and the energy is high, which
together suggest an ionized region.

The HRI observation demonstrates that the soft 
X-ray emission is extended to a radius of 4.8 kpc.
This extended emission accounts for half the observed counts, in
contrast to the small fraction of thermal soft X-ray flux detected
spectroscopically. 

We ascribe the discrepency of the image and spectrum to variability
of NGC 6221 in the two years between these observations.
The total flux, as well as the relative
fractions of resolved and thermal emission are significantly different 
between the \rosat{} and \asca{} data sets, 
with greater flux measured by \asca.  
The total observed HRI count rate is consistent with the \asca{} flux 
due to only two of the three soft components of the spectral model: 
the thermal and scattered components.  The
relative contributions of these two components correspond 
to the extended and unresolved components of the image.
The total flux measured with \asca{} then includes an additional component,
the weakly absorbed intrinsic AGN, which dominates the soft emission.

Flux variability may be due to a change in either the AGN itself 
or the intrinsic column density between the 
\rosat{} and \asca{} observations.    An increase in the AGN
power by a factor of about 5 between the observations is consistent
with the data.  Alternatively, $N_H$ may have diminished significantly.
The intrinsic column density we fit
to the 1997 data, $N_H= 1.1\times10^{22}\psc$, is extremely low.
If it had been more typical of the rest of the Sy2/SB sample,
$N_H \approx 10^{23}\psc$, at the time the \rosat{} data were obtained,
none of the \rosat-detected flux would have been due to the intrinsic 
AGN. 
NGC 6221 is the host of SN1990W, 
a type Ib \citep{Whe94} supernova, but this is unlikely to be
the source of X-ray variation or the extended emission.

\subsection{NGC 7130}
NGC 7130 is peculiar and contains a bar, detected at near-infrared wavelengths \citep{Mul97}.

The X-ray spectrum is best fit with a single
power law and thermal emission.  The latter dominates, accounting for
57\% of the soft X-ray flux.  In this model, 
the absorption on the power law is low,
consistent with the Galactic value alone, which implies that 
we do not observe the buried AGN directly.  The quality of these
\asca{} data is not extremely high, however.  
Although the complex model does not significantly improve the fit,
it yields reasonable results.
We find 
$N_H=10^{24}\psc$ and a scattered fraction $f_{scatt}=0.01$ in this
case, while the thermal contribution does not change.

We detect extended emission in the extremely short 
(5 ks) HRI exposure, with $R_{kpc}=16$.
\subsection{NGC 7582}
NGC 7582 is a group member and contains a bar. 

As mentioned above, medium energy emission from NGC 7582
varies \citep{Xue98}, so we have treated the two \asca{}
observations separately and combined each with the single
PSPC data set for the joint fits.
In both cases, the best-fitting model includes a scattered power law,
thermal emission, and a narrow line near 6.3 keV.
We attribute the spectral variation to a change in absorbing column density.
The intrinsic AGN is strong in NGC 7582, nearly 
$10^{-2} {\rm \,photons\,keV^{-1}\,cm^{-2}\,s^{-1}}$ at 1 keV.
The intrinsic AGN thus contributes to the soft X-ray flux,
so the thermal emission accounts for 0.16 and 0.28 of the total
in the two observations.

The X-ray emission is extended, on scales of 3.1 and 5.1 kpc in the
HRI and PSPC, respectively. The extended fraction is greater 
in the HRI observation (0.56), from 1995 May, than in 
the PSPC observation (0.38), taken in 1993 May.
We also observe complex structure in this nearby galaxy, and differences
between the hard and soft PSPC images.  

The optical spectrum of NGC 7582 has varied in recent years, exhibiting broad
lines as of July 1998. 
\citet{Are99} discuss several possible causes for this
transformation to an optical spectrum characteristic of a Seyfert 1 and
favor the appearance of a Type IIn supernova in the starburst.  These
and other X-ray observations \citep{Tur00} suggest, however, that the
column density that obscures the nucleus varies.  Although a simple
increase in extinction with the local reddening law is not consistent
with the optical continuum variations, the contribution of the
starburst component may not be adequately distinguished in the
present optical spectra \citep{Are99}.

Discussed above (\S\ref{subsec:ascaimage}), the nearby companion, RX J231829.9-422041, is 
much fainter than NGC 7582, especially at harder energies.
The spectrum of RX J231829.9-422041 is soft.
We fit the PSPC spectrum of this source alone
with a single power law, $\Gamma =2.1 (+0.44, -0.48)$ and Galactic absorption.

\subsection{NGC 7674}
NGC 7674 is interacting with UGC 12608, located 17 kpc ($32\arcsec$) toward
the northeast. 
This galaxy exhibits broad lines in polarized light \citep{Mil90,Tra95a}.

NGC 7674 was observed with the PSPC for only 3 ks. 
The best fit to these limited data
is a power law, with $\Gamma = 2.0 (+0.72, -1.3)$.
Although the measurement errors to not constrain the extended
fraction of soft X-ray emission well, this galaxy is resolved, 
with $R_{kpc}=45$.

\section{Conclusions\label{sec:concl}}
In this work, we present the results of X-ray imaging and spectroscopy
of a sample of Seyfert/starburst composite galaxies.  These are active
galaxies that also contain significant star formation.  The sample is
flux-limited, based on optical 
([\ion{O}{3}] $\lambda\lambda 4959+5007$) and radio (1.4 GHz) nuclear fluxes,
and the starbursts are detected in optical and UV observations.
Acknowledging the relevant physical processes of
these galaxies, we find spatial and spectral evidence for both
AGN and starburst activity in the sample members. 

The \rosat{} images of 85\% of this sample of Seyfert/starburst composite
galaxies reveal extended emission.  In several cases, where the PSPC
data are of extremely high quality, we also spatially differentiate soft and
hard emission in broad X-ray bands.  We expected that emission processes
other than the AGNs would be important in these galaxies, for they were
specifically selected to contain a circumnuclear starburst.  By analogy with
nearby starburst galaxies, we attribute the extended X-ray emission to 
thermal gas produced in a starburst-driven outflow.

We apply a consistent set of spectroscopic models to the X-ray emission,
detected primarily with \asca{} and supplemented by the \rosat{} PSPC.
All the \asca{} spectra require models that are more complex than a single
absorbed power law.  This model deficiency is most apparent as a soft X-ray excess,
a high excess of counts compared with extrapolation of the power law
at low energies, in the residuals.
Applying the progressively more complex models described above, we 
consistently identify the additional power law and thermal components
in the present sample.

The principal components of the resultant best-fitting models are
power laws,  emission lines, and thermal emission.  The power laws
represent the underlying AGN and its scattered emission, although in
many cases we do not detect the AGN directly because it is heavily
absorbed by material in the galaxy.  The emission lines, centered near
6.4 keV, are also due to the AGN.  In most of our sample members, the lines are
unresolved spectrally, which is typical in Sy 2s, where this emission
originates some distance from the nucleus. 
With one exception (NGC 6221), all the AGNs are heavily-absorbed,
with $N_H>10^{23}\psc$,
and the strength of the Fe K lines are consistent with these column densities.

Thermal emission dominating 
at energies less than about 1 keV is the X-ray
signature of the starburst.  An outflowing superwind, powered by the
stellar winds and supernovae of the starburst, are its immediate origin.
Only in one case (Mrk 477) do we fail to detect significantly this
starburst signature.

The imaging and spectroscopy results are generally consistent, supporting
the association of the spatially extended and spectrally-identified thermal
emission.  Thus, we detect the starburst component of these composite
galaxies in X-rays.

In Paper II, we apply the results obtained through the detailed 
data analysis of the present work and relate them to the larger
issues of the Seyfert-starburst connection.  
We compare the X-ray characteristics
of these Sy2/SB composites with ``pure'' starbursts and AGNs
and illustrate the composite nature of the Sy2/SBs at other wavelengths.
We also identify the distinctive signatures 
of the various energy sources in these different 
types of galaxies using multi-wavelength diagnostics.

\begin{acknowledgements} 
This work was supported by NASA grants  NAG5-6917 and NAG5-6400.
This research has made use of the NASA/IPAC Extragalactic Database
(NED) which is operated by the Jet Propulsion Laboratory,
California Institute of Technology, under contract with the
National Aeronautics and Space Administration,
the Astronomical Data Center at NASA Goddard Space Flight Center,
and the High Energy Astrophysics Science Archive Research Center Online Service
provided by the NASA Goddard Space Flight Center. 
The Digitized Sky Surveys were produced at the 
Space Telescope Science Institute under U.S. 
Government grant NAG W-2166. The images of these surveys are 
based on photographic data obtained using the Oschin Schmidt 
Telescope on Palomar Mountain and the UK
Schmidt Telescope.

\end{acknowledgements}

\rotate

\begin{deluxetable}{lllllllrrcclc}
\tablewidth{0pt}
\tablecaption{Basic Properties of the Selected Galaxies\label{tab:info}}
\tablehead{
\colhead{Galaxy}&\multicolumn{3}{c}{RA}&\multicolumn{3}{c}{DEC}&\colhead{D} &\colhead{Scale}
&\colhead{Galactic $N_H$}
&\colhead{$i$}& 
\colhead{References}\\
&\colhead{(h}&\colhead{m}&\colhead{s)}&\colhead{($^\circ$}&\colhead{\arcmin}&\colhead{\arcsec)}
&\colhead{(Mpc)}&\colhead{(pc/$\arcsec$)}&\colhead{($10^{20}{\rm \,cm^{-2}}$)}
&\colhead{($^\circ$)}
}
\startdata
NGC 1068  & 02& 42& 40.71   & $-00$ & 00&  47.8 & 15.2&  74 & 3.0 & 32 &1, 4\\   
Mrk 1066  & 02 & 59 & 58.59 & $+36$ & 49 & 14.3 & 48  & 233 & 12  & 54 &2, 4, 6 \\
Mrk 1073  & 03 & 15 & 01.43 & $+42$ & 02 & 09.4 & 93  & 452 & 1.1 & 42 &2, 5, 6 \\
Mrk 78    & 07 & 42 & 41.73 & $+65$ & 10 & 37.5 & 149 & 720 & 4.1 & 60 &2, 4, 6\\
IC 3639   & 12 & 40 & 52.88 & $-36$ & 45 & 21.5 & 44  & 212 & 5.1 & 34 &2, 5, 7\\
NGC 5135  & 13 & 25 & 43.97 & $-29$ & 50 & 02.3 & 55  & 266 & 4.7 & 46 &2, 4, 7, 8, 9\\
Mrk 266   & 13 & 38 & 17.69 & $+48$ & 16 & 33.9 & 111 & 540 & 1.5 & \nodata & 1, 4, 10\\
Mrk 273   & 13 & 44 & 42.11 & $+55$ & 53 & 12.6 & 151 & 732 & 0.97& \nodata & 1, 5, 6\\
Mrk 463   & 13 & 56 & 02.87 & $+18$ & 22 & 19.5 & 199 & 963 & 2.1 & 10 & 2, 5, 6\\
Mrk 477   & 14 & 40 & 38.11 & $+53$ & 30 & 16.0 & 151 & 733 & 1.3 &\nodata & 2, 5, 7, 11\\
NGC 6221  & 16 & 52 & 46.67 & $-59$ & 12 & 59.0  & 20  &  96 & 15 & 43  & 2, 5, 8\\
NGC 7130  & 21 & 48 & 19.48 & $-34$ & 57 & 09.2 & 65  & 313 & 2.0 & 29 & 2, 5, 7, 8, 9\\
NGC 7582  & 23 & 18 & 23.50 & $-42$ & 22 & 14.0 & 21  & 102 & 1.5 & 62 &3, 5, 8, 9 \\
NGC 7674  & 23 & 27 & 56.72 & $+08$ & 46 & 44.5 & 116 & 563 & 5.2 & 25 &2, 4, 6\\
\enddata

\tablerefs{
(1) Murphy et al. (1996); 
(2) Dickey \& Lockman (1990); 
(3) Elvis, Lockman, \& Wilkes (1989); 
(4) Schmitt et al. (1997); 
(5) Whittle (1992);        
(6) Gonzalez-Delgado et al. (2000); 
(7) Gonzalez-Delgado et al. (1998); 
(8) Cid Fernandes et al. (1998); 
(9) Schmitt et al. (1999); 
(10) Wang et al. (1997); 
(11) Heckman et al. (1997) 
}
\end{deluxetable}

\begin{deluxetable}{llrrllrrllrrc}
\tabletypesize{\small}
\tablewidth{0pt}
\tablecaption{Seyfert/Starburst X-ray Observations\label{tab:obs}}
\tablehead{
&\multicolumn{3}{c}{HRI}
&&\multicolumn{3}{c}{PSPC}&&\multicolumn{4}{c}{ASCA}\\
\cline{2-4} \cline{6-8} \cline{10-13} \\
\colhead{Galaxy}
&\colhead{Date}&\colhead{Exposure}&\colhead{Rate}
&&\colhead{Date}&\colhead{Exposure}&\colhead{Rate}
&&\colhead{Date}&\colhead{Exposure\tablenotemark{a}}
&\colhead{Rate\tablenotemark{a}}&\colhead{SIS CCD Mode}\\
&&\colhead{(ks)}&\colhead{(cts s$^{-1}$)}
&&&\colhead{(ks)}&\colhead{(cts s$^{-1}$)}
&&&\colhead{(ks)}&\colhead{(cts s$^{-1}$)}
}
\startdata

NGC 1068  & 1995 Jul & 71.4 & 0.638 &&1993 Jul & 5.30 & 1.562 &&  1996 Aug & 93.8 & 0.372 & 1    \\
Mrk 1066  & 1996 Feb & 15.4 & 0.004 &&\nodata&\nodata&\nodata &&  1997 Aug & 34.3 & 0.010 & 1    \\
          & 1998 Jan & 25.6 & 0.004 &&\nodata&\nodata&\nodata &&   \nodata&\nodata&\nodata       \\
Mrk 1073  & 1998 Aug & 20.7 & 0.004 &&\nodata&\nodata&\nodata &&  1999 Sep & 25.7 & 0.012 & 1    \\
Mrk 78    & 1996 Mar & 16.7 & 0.004 &&1991 Mar & 14.3 & 0.008 && \nodata&\nodata&\nodata&\nodata \\
IC 3639   & 1998 Jan & 39.1 & 0.005 &&\nodata&\nodata&\nodata &&  1998 Jul & 37.0 & 0.008 & 1    \\
NGC 5135  & 1998 Jan & 35.0 & 0.014 &&1992 Jul & 9.24 & 0.040 &&  1995 Jan & 44.4 & 0.015 & 1    \\
Mrk 266   & 1996 Jun & 39.4 & 0.012 &&1991 Dec & 9.54 & 0.032 &&  1999 May & 18.0 & 0.013 & 1    \\
Mrk 273   & 1992 May & 19.2 & 0.004 &&1992 Jun & 19.9 & 0.010 &&  1994 Dec & 40.6 & 0.009 & 1    \\
Mrk 463   & 1992 Jan & 1.05 &\nodata&&1992 Jan & 11.1 & 0.017 &&  1994 Jan & 36.6 & 0.007 & 2    \\
Mrk 477   & 1997 Dec & 20.5 & 0.008 &&\nodata&\nodata&\nodata &&  1995 Dec & 26.7 & 0.012 & 1    \\
          &\nodata  &\nodata&\nodata&&\nodata&\nodata&\nodata &&  1995 Dec & 17.1 & 0.012 & 1    \\
NGC 6221  & 1995 Sep & 11.9 & 0.033 &&\nodata&\nodata&\nodata &&  1997 Sep & 35.1 & 0.315 & 1    \\
NGC 7130  & 1998 Apr & 5.26 & 0.010 &&\nodata&\nodata&\nodata &&  1996 Nov & 35.9 & 0.013 & 1    \\
NGC 7582  & 1995 May & 17.3 & 0.015 &&1993 May & 7.24 & 0.042 &&  1994 Nov & 20.2 & 0.087 & 2    \\
          &\nodata  &\nodata&\nodata&&\nodata&\nodata&\nodata &&  1996 Nov & 37.1 & 0.115 & 1    \\
NGC 7674  & 1992 Jun & 0.89 &\nodata&&1992 Dec & 3.71 & 0.016 && \nodata&\nodata&\nodata&\nodata \\
\enddata 
\tablenotetext{a}{SIS0}
\tablecomments{
Rates are background-subtracted. 
}
\end{deluxetable}

\begin{deluxetable}{lrllcrllcl}
\tablewidth{0pt}
\tablecaption{Extended and Thermal Soft X-ray Emission\label{tab:extend}}
\tablehead{
&\multicolumn{3}{c}{HRI}&&\multicolumn{3}{c}{PSPC}&&\multicolumn{1}{c}{ASCA}\\
\cline{2-4}\cline{6-8}\cline{10-10}\\ 
\colhead{Galaxy}&\colhead{$R$\tablenotemark{a}$_{kpc}$}&\colhead{$f$\tablenotemark{b}}
&\colhead{$f$\tablenotemark{c}}
&&\colhead{$R$\tablenotemark{a}$_{kpc}$}&\colhead{$f$\tablenotemark{b}}
&\colhead{$f$\tablenotemark{c}}&&\colhead{$f$\tablenotemark{d}$_{therm}$}
}
\startdata
NGC 1068  &  7.4 & $0.51^{+0.05}_{-0.09}$& $0.73 \pm 0.02$      &&  7.4    & $ 0.40   \pm0.02        $ & $ 0.60\pm 0.06$       &&  0.80  \\
Mrk 1066  &  0.0 & $<0.15               $& $<0.71        $      && \nodata &  \nodata                  & \nodata               &&  0.51  \\
Mrk 1073  &  9.5 & $0.53^{+0.18}_{-0.28}$&$0.84^{+0.16}_{-0.50}$&& \nodata &  \nodata                  & \nodata               &&  0.91  \\
Mrk 78    &  17  & $0.38^{+0.20}_{-0.30}$&$0.72^{+0.28}_{-0.54}$&& 36      & $ <0.62                 $ & $0.72^{+0.28}_{-0.45}$&&\nodata  \\
IC 3639   &  6.8 & $0.39^{+0.11}_{-0.21}$& $0.68 \pm 0.31$      && \nodata &  \nodata                  & \nodata               &&  1.00  \\
NGC 5135  &  5.3 & $0.46^{+0.06}_{-0.16}$& $0.65 \pm 0.20$      &&  13     & $ 0.19  ^{+0.07}_{-0.11}$ & $ < 0.32      $       &&  0.54  \\
Mrk 266   &  22  & $0.78^{+0.06}_{-0.16}$&$0.92^{+0.08}_{-0.20}$&&  32     & $ 0.49  ^{+0.08}_{-0.12}$ & $0.77^{+0.23}_{-0.28}$&&  0.43  \\
Mrk 273   &  18  & $<0.39               $&$0.53^{+0.47}_{-0.49}$&& 37      & $<0.28                  $ & $ < 0.62      $       &&  0.43  \\
Mrk 463   & \nodata            & \nodata &   \nodata            && 48      & $<0.41                  $ & $ < 0.67      $       &&  0.31  \\
Mrk 477   &  15  & $<0.30               $& $0.54 \pm 0.35$      && \nodata &  \nodata                   & \nodata              &&  0.00  \\
NGC 6221  &  4.8 & $0.52^{+0.07}_{-0.17}$&$0.96^{+0.04}_{-0.22}$&& \nodata &  \nodata                   & \nodata              &&  0.05  \\
NGC 7130  &  16  & $0.44^{+0.23}_{-0.33}$& $< 0.63       $      && \nodata &  \nodata                   & \nodata              &&  0.57  \\
NGC 7582  &  3.1 & $0.56^{+0.09}_{-0.19}$&$0.94^{+0.06}_{-0.27}$&&  5.1    & $ 0.38  ^{+0.08}_{-0.12}$ & $ 0.54\pm 0.28$       &&0.16, 0.28  \\
NGC 7674  & \nodata            & \nodata &   \nodata            && 45      & $0.65   ^{+0.48}_{-0.52}$ & $ < 0.57      $       &&\nodata  \\
\enddata
\tablenotetext{a}{Maximum radius of extended emission (kpc).}
\tablenotetext{b}{Minimum extended fraction of total emission, with 90\% confidence errors.}
\tablenotetext{c}{Constrained measurement of extended fraction of total emission.} 
\tablenotetext{d}{Thermal fraction of soft emission in spectrum.} 
\end{deluxetable}

\begin{deluxetable}{llllllc}
\tabletypesize{\footnotesize}
\tablewidth{0pt}
\tablecaption{Power Law Models\label{tab:pl}}
\tablehead{
\colhead{Galaxy}&\colhead{$N$\tablenotemark{a}$_H$}&\colhead{$\Gamma$\tablenotemark{b}}&\colhead{$A1$\tablenotemark{c}}
&\colhead{$F$\tablenotemark{d}$_{2-10}$}&\colhead{$F$\tablenotemark{e}$_{0.5-2}$}&\colhead{$\chi^2/$dof}
}
 
\startdata
\sidehead{{\it ASCA}}
NGC 1068 &$3:^{+0.1}_{-0:}$&$3.2\pm0.02$&$40\pm1.0  $&$20 \pm0.5 $&$ 86\pm2.0 $& 8533/813 \\
Mrk 1066 &$  12:^{+  9.4}_{-0:}$&$1.8\pm0.25$&$0.91^{+0.22}_{-0.17}$&$2.9^{+0.45}_{-0.41}$&$1.4^{+0.22}_{-0.20}$&  100/76 \\
Mrk 1073 &$1.4:^{+13}_{-0:}$&$1.2^{+0.38}_{-0.40}$&$0.61^{+0.26}_{-0.23}$&$5.2\pm1.0$&$1.3\pm0.26$&132/64\\
IC 3639 &$5.0:^{+20}_{-0:}$&$1.3^\pm0.4$&$0.51^{+0.3}_{-0.2}$&$3.7^{+2.1}_{-1.6}$&$ 1.0^{+0.6}_{-0.4}$&  64/54 \\ 
NGC 5135 &$4.7:^{+4.4}_{-0:}$&$3.0^{+0.3}_{-0.2}$&$1.9\pm0.4$& $1.4\pm0.3$& $5.5\pm1.0$& 215/100 \\
Mrk 266 & $1.5:^{+10}_{-0:}$&$2.2^{+0.8}_{-0.6}$&$1.1^{+0.7}_{-0.4}$&$2.0^{+1.0}_{-0.5}$&$2.2^{+1.1}_{-0.6}$&28/34\\
Mrk 273 &$ 1.0:^{+5.1}_{-0:}$&$1.3^{+0.31}_{-0.32}$&$0.46^{+0.16}_{-0.14}$&$3.5^{+0.70}_{-0.46}$&$1.0^{+0.2}_{-0.13}$ &123/61 \\
Mrk 463 &$  2.0:^{+7.5}_{-0:}$&$0.65^{+0.24}_{-0.23}$&$ 0.27^{+0.094}_{-0.077}$& $6.1\pm1.2$ & $0.64\pm0.13$ & 123/72 \\
Mrk 477   &$1.3:^{+10}_{-0:}$ &$ 0.12\pm0.3$&$ 0.22^{+0.2}_{-0.1}$&$ 13\pm1.4$&$0.62\pm0.06  $&65/106\\
NGC 6221  &$78^{+4.9}_{-4.7}$&$1.7\pm0.05$&$38^{+2.7}_{-2.5}$&$140\pm7.0$&$24\pm1.2$&846/769\\
NGC 7130 &$  2.0:^{+20}_{-0:}$&$2.6^{+0.7}_{-0.3}$&$ 1.2^{+0.7}_{-0.4}$&$2.4^{+1.6}_{-0.8}$&$ 2.4^{+ 1.6}_{- 0.8}$& 103/75 \\
NGC 7582 (1994) &$ 520^{+120}_{-70}$&$1.0^{+ 0.3}_{-0.2}$&$ 16^{+9.8}_{-4.1}$&$160\pm7.0$ & $34^{+1.5}_{-1.4}$ &527/286 \\
NGC 7582 (1996) &$760^{+72}_{-90}$&$ 1.1^{+0.1}_{-0.2}$&$21\pm7.5$&$160\pm32$ & $0.26\pm0.1$    &1614/469 \\
 
\sidehead{{\it ASCA + PSPC}}
NGC 1068 &$4.6\pm0.2$&$3.3\pm0.03  $&$42\pm11  $&$19\pm0.4 $&$ 85\pm1.6 $& 8975/936 \\
NGC 5135 &$8.0^{+2.3}_{-1.6}$&$3.2^{+0.3}_{-0.2}$&$2.1^{+0.5}_{-0.4}$&$1.0\pm0.2$&$3.6\pm0.6$&242/121\\
Mrk 266 & $3.7^{+2.4}_{-1.5}$& $2.7^{+0.6}_{-0.5}$&$1.2^{+0.3}_{-0.2}$& $1.2\pm0.2$&$2.4\pm0.4$& 64/51\\
Mrk 273 &$ 2.2^{+2.1}_{-1.2}$&$1.4^{+0.24}_{-0.22}$&$0.51^{+0.085}_{-0.11}$& $3.4^{+0.41}_{-0.58}$ &$1.0^{+0.12}_{-0.17}$  & 151/72 \\
Mrk 463 &$  2.0:^{+1.8}_{-0:}$&$0.78^{+0.23}_{-0.22}$&$0.30^{+0.076}_{-0.066}$&$5.6^{+1.1}_{-0.95}
$&$0.65^{+1.2}_{-1.1}$& 168/89 \\
NGC 7582 (1994) &$  520^{+80}_{-110}$&$1.1^{+0.2}_{-0.3}$&$18^{+6.9}_{-6.7}$& $160^{+7.1}_{-6.9}$& $0.77^{+0.04}_{-0.03}$   &795/302 \\
NGC 7582 (1996) &$740\pm80$&$1.1\pm0.1$&$20\pm7.8$&$160\pm32$ & $0.26\pm0.1$ &1861/484 \\
 
\enddata
 
\tablenotetext{a}{Column density in units of $10^{20}{\rm\,cm^{-2}}$.}
\tablenotetext{b}{Photon index of power law.}
\tablenotetext{c}{Normalization of power law in units of
$10^{-4} {\rm\,photons\, keV^{-1}\,cm^{-2}\,s^{-1}}$ at 1 keV.}
\tablenotetext{d}{2.0--10.0 keV model flux in  SIS0 detector in units of
$10^{-13}{\rm\,erg\,cm^{-2}\,s^{-1}}$.}
\tablenotetext{e}{0.5--2.0 keV model flux in SIS0 detector in units of
$10^{-13}{\rm\,erg\,cm^{-2}\,s^{-1}}$.}
\tablecomments{
Errors are 90\% confidence limits for two interesting parameters, except
fluxes, where errors are 90\% confidence for one parameter.
Parameters that are constrained by hard limits are marked with a colon.}
 
\end{deluxetable}

\begin{deluxetable}{llllcllc}
\tabletypesize{\small}
\tablewidth{0pt}
\tablecaption{Scattered Power Law Models ($\Gamma=1.9$)\label{tab:pl2}}
\tablehead{
&\multicolumn{3}{c}{Hard Component}&&\multicolumn{2}{c}{Soft Component}\\
\cline{2-4} \cline{6-7}\\
\colhead{Galaxy}&\colhead{$N_H$}&\colhead{$A1$} 
&\colhead{$F_{2-10}$}&&\colhead{$A2$\tablenotemark{a}}
&\colhead{$F_{0.5-2}$}
&\colhead{$\chi^2/$dof}
}
\startdata
\sidehead{{\it ASCA}}
NGC 1068 &$3:^{+2}_{-0:} $&$20^{+0.5}_{-5.7}$&$59^{+1.2}_{-18} $&&$0:^{+64}_{-0:}  $&$40.0^{+0.8}_{-12} $& 24682/813 \\
Mrk 1066 &$17100^{+32000 }_{-5600}$&$38^{+180}_{-34}$&$6.1\pm0.52$&&$0.94\pm0.10 $&$ 1.5\pm0.13$& 85/76  \\
Mrk 1073 & $11500^{+15000}_{-4200}$& $32^{+89}_{-26}$& $9.3\pm1.4$&&$0.98^{+0.19}_{-0.20}$&$2.0\pm0.30$&118/64\\
IC 3639 &$ 3200^{+4600}_{-2000}$&$3.4^{+6.0}_{-2.1}$&$5.1^{+1.8}_{-1.7}$&&$0.73\pm0.25$&$1.4\pm0.2$& 52/54 \\
NGC 5135 &$ 22700^{+38000}_{-8900}$&$ 120^{+20000}_{-110}$&$8.3^{+2.5}_{-2.4}$&&$1.1^{+0.24}_{-0.17}$&$ 2.1^{+0.2}_{-0.1}$& 310/100   \\
Mrk 266 &$3300^{+1E+08:}_{-3300:}$& $2.8^{+22}_{-2.8:}$&$4.9\pm1.2$&&$0.89^{+0.31}_{-0.32}$&$1.8\pm0.5$&25/34\\
Mrk 273 &$ 5300^{+3600}_{-1800}$&$7.9^{+7.4}_{-3.8}$&$6.3\pm1.3$ &&$ 0.67^{+ 0.068}_{-0.12}$& $1.4^{+0.095}_{-0.19}$  & 68/61  \\
Mrk 463 &$ 3300^{+1700}_{-1200}$&$6.5^{+6.7}_{-3.1}$&$6.8\pm1.4$&&$0.52^{+0.060}_{-0.095}$&$1.0^{+0.080}_{-0.15}$& 88/72  \\
Mrk 477  &$3220^{+1900}_{-1200}$&$13^{+7.6}_{-5.9}$&$13\pm2.0$&&$0.62^{+0.21}_{-0.22}$&$1.2^{+0.1}_{-0.2}$&41/106\\
NGC 6221 &$110^{+7.0}_{-6.7}$&$45\pm1.5$&$130\pm2.7$&&$4.3^{+1.0}_{-1.1}$&$24\pm0.5$&846/769\\
NGC 7130 &$36000^{+1E+08:}_{-36000:}$&$330^{+4700}_{-330:}$&$5.3^{+7.0}_{-5.0}$&&$ 0.86^{+0.35}_{-0.22}$&$1.9^{+ 0.3}_{-0.04}$& 126/75 \\
NGC 7582 (1994) &$1100^{+80}_{-70}$&$87^{+6.3}_{-6.0}$&$ 140\pm6.3$&&$ 2.1^{+0.29}_{-0.33}$&$4.5^{+0.4}_{-0.6}$& 357/286 \\
NGC 7582 (1996)&$ 1500\pm50 $&$ 97^{+4.6}_{-4.5}$&$ 140\pm4.2$&&$2.1^{+0.14}_{-0.17}$&$4.4^{+0.2}_{-0.3}$& 552/469 \\
\sidehead{{\it ASCA + PSPC}}
NGC 1068 &$3:^{+4E-03}_{-0:}$&$20^{+0.5}_{-2.3}$&$59^{+1.1}_{-3.2} $&&$0.0:^{+2.6}_{-0:}  $&$40^{+0.7}_{-2.2} $&  27845/936\\
NGC 5135 &$ 22600^{+34000}_{-7300}$&$ 120^{+260}_{-110}$&$8.2^{+ 2.9}_{-2.4}$&&$1.1^{+0.22}_{-0.18}$&$ 2.1^{+0.3}_{-0.2}$& 368/121 \\
Mrk 266 &$3470^{+1E+08:}_{-3500:}$&$2.6^{+42}_{-2.6:}$&$4.6\pm0.6$&&$0.90^{+0.12}_{-0.13}$&$1.8\pm0.2$& 72/51\\
Mrk 273 &$ 5300^{+3500}_{-1600}$&$7.8^{+7.5}_{-3.7}$&$6.2\pm1.2$ &&$ 0.64^{+ 0.079}_{-0.073}$& $1.3^{+0.12}_{-0.25}$  & 131/72  \\
Mrk 463 &$ 3300^{+1700}_{-990}$&$ 6.5^{+6.0}_{-3.2}$&$ 6.8\pm1.4$&&$0.50^{+0.069}_{-0.059}$&$ 0.95^{+0.10}_{-0.084}$& 109/89  \\
NGC 7582 (1994) &$ 1100^{+80}_{-70}$&$87^{+6.2}_{-6.0}$ &$140^{+1.8}_{-10}$&&$2.1^{+0.27}_{-0.21}$&$ 4.0^{+0.4}_{-0.3}$& 397/300   \\ 
NGC 7582 (1996)&$ 1500^{+57}_{-55}$&$97^{+3.4}_{-3.3}$&$ 140\pm2.2 $&&$ 2.1^{+0.12}_{-0.15}$&$4.2^{+0.15}_{-0.20}$& 594/483   \\ 
\enddata

\tablenotetext{a}{Normalization of the soft power-law component in units of 
$10^{-4} {\rm\,photons\, keV^{-1}\,cm^{-2}\,s^{-1}}$ at 1 keV.}
\tablecomments{
Power law photon index is fixed at 1.9.
Other symbols and errors are described in Notes to Table \ref{tab:pl}}
\end{deluxetable}

\begin{deluxetable}{llllclllc}
\tabletypesize{\small}
\tablewidth{0pt}
\tablecaption{Power Law + Thermal Models ($\Gamma=1.9$)\label{tab:plm}}

\tablehead{
&\multicolumn{3}{c}{Hard Component}&&\multicolumn{3}{c}{Soft Component}\\
\cline{2-4} \cline{6-8}\\
\colhead{Galaxy}&\colhead{$N_H$}&\colhead{$A1$} 
&\colhead{$F_{2-10}$}&&\colhead{$kT$\tablenotemark{a}}&\colhead{$A3$\tablenotemark{b}}&\colhead{$F_{0.5-2}$}
&\colhead{$\chi^2/$dof}
}
\startdata
\sidehead{{\it ASCA}}
NGC 1068 &$3:^{+0.2}_{-0:} $&$ 12\pm0.3 $&$37^{+0.6}_{-0.8}$&&$0.68\pm0.01 $&$19^{+0.50}_{-0.55} $&$67^{+1.2}_{-1.4} $& 4214/812 \\
Mrk 1066 & $75^{+69}_{-46}$& $1.1^{+0.23}_{-0.29}$ & $3.0^{+0.48}_{-0.65}$&&$0.88^{+0.23}_{-0.24}$&$0.45\pm0.26$&$1.3^{+0.21}_{-0.29}$&90/75\\
Mrk 1073 & $400^{+390}_{-220}$&$2.2^{+0.89}_{-0.70}$&$4.7^{+0.74}_{-0.85}$&&$1.0^{+0.47}_{-0.29}$&$0.70^{+0.82}_{-0.28}$&$1.3\pm0.26$&119/63\\
IC 3639 &$2100^{+3000}_{-1300}$&$3.4^{+ 3.2}_{-1.7}$&$ 4.8\pm1.2$&&$ 2.3^{+2.9}_{-0.7}$&$1.7\pm0.53$&$1.2\pm0.29$& 48/53 \\
NGC 5135 &$ 4.7:^{+20}_{-0:}$&$ 0.71\pm0.2$&$2.2\pm0.2$&&$ 0.75^{+0.09}_{-0.1}$&$0.78\pm0.19$&$ 2.9\pm0.3$& 154/99 \\
Mrk 266 & $160^{+330}_{-150}$&$1.2^{+1.1}_{-0.6}$&$3.1\pm1.1$&&$0.78\pm0.3$&$0.73^{+0.35}_{-0.39}$&$1.8\pm0.6$&13/33\\
Mrk 273 &$ 4200^{+2900}_{-1500}$&$ 6.6^{+5.5}_{-2.4}$&$ 5.8\pm1.2$&&$ 3.2^{+ 2.2}_{-1.0}$&$ 1.7^{+0.27}_{-0.30}$&$ 1.2^{+0.14}_{-0.16}$& 68/60  \\
Mrk 463 &$ 3400^{+3500}_{-730}$&$ 6.8^{+8.8}_{-3.5}$&$ 7.1\pm1.4$&&$ 5.7^{+24}_{-2.9}$&$ 1.5^{+0.70}_{-0.25}$&$ 0.84^{+0.17}_{-0.11}$& 88/71 \\
Mrk 477 &$3700^{+3800}_{-1600}$&$14^{+20}_{-6.5}$&$13\pm2.3$ && $9.0^{+91:}_{-6.2}$&$1.9^{+2.1}_{-0.70}$&$1.1\pm0.2$&41/105\\
NGC 6221&$100^{+7.6}_{-4.5}$&$39^{+2.6}_{-2.5}$&$140\pm6.7$&&$74^{+26:}_{-62}$&$25^{+6.5}_{-8.2}$&$24\pm1.2$&832/768\\
NGC 7130 &$ 16^{+62}_{-14:}$&$ 0.60\pm0.3$&$ 1.8\pm0.3$&&$ 0.79^{+0.2 }_{-0.1}$&$ 0.56^{+0.29}_{-0.25}$&$ 2.0^{+0.3}_{-0.4}$& 74/74 \\
NGC 7582 (1994) &$ 1130^{+50}_{-90}$&$ 87^{+6.6}_{-6.9}$&$ 140\pm6.5 $&&$ 13^{+87:}_{-8.9}$&$ 7.1^{+5.0}_{-1.7}$&$ 3.9^{+0.5}_{-0.4}$& 360/285   \\
NGC 7582 (1996) &$ 1480^{+70}_{-60}$&$ 97^{+4.7}_{-4.6}$&$ 140\pm4.3$&&$ 6.4^{+5.6}_{-2.0}$&$ 6.3^{+0.79}_{-0.64}$&$ 3.9\pm0.2$& 572/468 \\
\sidehead{{\it ASCA + PSPC}}
NGC 1068 &$3:^{+0.01}_{-0:} $&$ 12\pm0.3 $&$38\pm0.7$&&$0.67\pm0.01 $&$19^{+0.53}_{-0.51} $&$67\pm1.3 $&7361/935 \\
NGC 5135 &$ 4.7:^{+1.6}_{-0:}$&$ 0.74\pm0.2$&$ 2.2\pm0.2$&&$ 0.73^{+0.1}_{-0.2}$&$ 0.75^{+0.24}_{-0.21}$&$ 2.9\pm0.3$& 194/120 \\
Mrk 266 &$1.5:^{+1.1}_{-0:}$&$0.75\pm0.2$&$2.1^{+0.4}_{-0.8}$&&$0.28^{+0.1}_{-0.05}$& $0.72^{+0.42}_{-0.33}$&$2.6^{+0.5}_{-0.9}$&43/50\\
Mrk 273 &$ 4000^{+3100}_{-1500}$&$ 6.4^{+5.9}_{-2.5}$&$5.7\pm1.1$ &&$ 2.9^{+2.1}_{-0.83}$&$1.7^{+0.27}_{-0.32}$& $1.2^{+0.12}_{-0.15}$  & 96/71  \\
Mrk 463 &$ 3100^{+2600}_{-1200}$&$ 5.9^{+6.6}_{-2.6}$&$ 6.5\pm1.3$&&$ 4.3^{+7.4}_{-2.0}$&$ 1.4^{+0.25}_{-0.28}$&$ 0.85^{+0.11}_{-0.13}$& 119/88 \\
NGC 7582 (1994) &$ 940\pm60$&$83^{+5.8}_{-5.6}$&$ 140\pm5.9$ &&$ 1.0^{+0.1}_{-0.2}$&$1.6^{+0.34}_{-0.36} $&$ 3.1\pm0.4$& 455/299  \\ 
NGC 7582 (1996)&$ 1300^{+46}_{-48}$&$ 94\pm3.2$&$140\pm2.1$ &&$ 0.56^{+0.059}_{-0.047}$&$ 17^{+3.7}_{-3.5}$& $3.4^{+0.28}_{-0.26}$  & 700/482  \\ 

\enddata

\tablenotetext{a}{Temperature of thermal plasma in keV.}
\tablenotetext{b}{Normalization of thermal component in units of $10^{-4}\times K$, where
$K=(10^{-14}/(4\pi D^2))\int n_e n_H dV, D$ is the distance to the source (cm), $n_e$
is the electron density (${\rm cm^{-3}}$), and $n_H$ is the hydrogen density (${\rm cm^{-2}}$).}

\tablecomments{
Power law photon index is fixed at 1.9.
Other symbols and errors are described in Notes to Table \ref{tab:pl}.
%and \ref{tab:pl2}. 
}

\end{deluxetable}

\begin{deluxetable}{llllrllllc}
\tablewidth{0pt}
\tablecaption{Scattered Power Law + Thermal Models ($\Gamma=1.9$)\label{tab:pl2m}}
\tablehead{
&\multicolumn{3}{c}{Hard Component}&&\multicolumn{4}{c}{Soft Component}\\
\cline{2-4} \cline{6-9}\\
\colhead{Galaxy}&\colhead{$N_H$}&\colhead{$A1$} 
&\colhead{$F_{2-10}$}&&\colhead{$A2$}&\colhead{$kT$}&\colhead{$A3$}
&\colhead{$F_{0.5-2}$}
&\colhead{$\chi^2/$dof}
}
\startdata

\sidehead{{\it ASCA}}
NGC 1068 &$9900^{+1000}_{-970} $&$ 86^{+18}_{-15} $&$58 \pm 3.3 $&&$11^{+0.28}_{-0.29} $&$0.69^{+0.01}_{-0.003}$ &$2.0\pm0.53 $&$ 66\pm2.3 $&  3363/811 \\
Mrk 1066 & $14700^{+28500}_{-9000}$&$26^{+2000}_{-23}$&$5.6^{+0.67}_{-0.78}$&&$0.82^{+0.13}_{-0.15}$&$0.90^{+0.90}_{-0.66}$&$0.17^{+0.28}_{-0.15}$&$1.5^{+0.18}_{-0.21}$& 79/74\\
NGC 5135 &$ 12900^{+16000}_{-6800}$&$ 25.7^{+120}_{-18}$&$ 6.2^{+1.5}_{-1.6}$&&$ 0.67^{+0.09}_{-0.15}$&$ 0.77^{+0.04 }_{-0.1}$&$ 0.79^{+0.28}_{-0.14}$&$ 2.8^{+0.5}_{-0.2}$& 108/98 \\
Mrk 273 &$ 4500^{+2200}_{-1200}$&$ 6.7^{+5.3}_{-3.0}$&$ 5.6\pm1.1$&&$ 0.49^{+0.16}_{-0.33}$&$ 1.1^{+1.8}_{-0.89}$&$ 0.20^{+0.28}_{-0.19}$&$ 1.2\pm0.24$& 62/59  \\
Mrk 463 &$ 2900^{+1700}_{-1200}$&$ 6.0^{+4.7}_{-2.8}$&$ 6.6\pm1.3$&&$ 0.40^{+0.080}_{-0.086}$&$ 0.71^{+0.53}_{-0.45}$&$ 0.15^{+0.080}_{-0.11}$&$ 1.1\pm0.17$& 80/70 \\
\sidehead{{\it ASCA + PSPC}}
NGC 1068 &$3:^{+10}_{-0:} $&$ 12^{+0.3}_{-1.7} $&$40^{+0.7}_{-0.6}$&&$0:^{+5.8}_{-0:} $&$0.67\pm0.01$ &$1.9^{+5.2}_{-5.1} $&$ 66^{+1.3}_{-1.2} $&  3363/811 \\
NGC 5135 &$ 14000^{+16000}_{-3700}$&$ 29^{+100}_{-21}$&$ 6.3\pm0.63$&&$ 0.71^{+0.11}_{-0.10}$&$ 0.68^{+0.14}_{-0.06}$&$ 0.79^{+0.10}_{-0.11}$&$ 2.9\pm0.29$& 150/119 \\
Mrk 273 &$ 3600^{+2000}_{-1100}$&$ 5.3^{+3.7}_{-2.0}$&$ 5.3\pm0.89 $&&$ 0.38^{+0.085}_{-0.097}$&$ 0.83^{+0.49}_{-0.26}$&$ 0.28^{+0.090}_{-0.097}$&$ 1.3^{+0.21}_{-0.23}$& 98/70\\
Mrk 463 &$ 2800^{+1300}_{-850}$&$ 5.8^{+5.0}_{-2.7}$&$ 6.6\pm1.3$&&$ 0.40^{+0.085}_{-0.094}$&$ 0.66^{+0.35}_{-0.39}$&$ 0.16^{+0.086}_{-0.084}$&$ 1.1\pm0.17$& 96/87  \\
NGC 7582 (1994) &$ 1100^{+84}_{-76}$&$ 86^{+5.4}_{-5.0}$&$ 140^{+4.1}_{-4.2}$&&$ 1.9^{+0.26}_{-0.52}$&$ 0.37^{+0.30}_{-0.14}$&$0.63^{+0.64}_{-0.35}$&$ 4.7^{+0.37}_{-0.80}$& 377/298  \\ 
NGC 7582 (1996) &$ 1400^{+58}_{-51}$&$ 96^{+3.4}_{-3.3}$&$ 150^{+2.4}_{-2.6}$&&$ 1.9{+0.20}_{-0.16}$&$ 0.59^{+0.25}_{-0.30}$&$ 0.27^{+0.18}_{-0.13}$&$ 4.3^{+0.23}_{-0.22}$& 575/481  \\ 
\enddata

\tablecomments{
Power law photon index is fixed at 1.9.
Symbols and errors are described in Notes to Tables \ref{tab:pl}--\ref{tab:plm}. 
}

\end{deluxetable}

\begin{deluxetable}{lllllllclllllc}
\tabletypesize{\scriptsize}
\tablewidth{0pt}
\tablecaption{Best-Fitting Models ($\Gamma=1.9$)\label{tab:best}}
\tablehead{
&\multicolumn{6}{c}{Hard Component}&&\multicolumn{5}{c}{Soft Component}\\
\cline{2-7} \cline{9-13}\\
\colhead{Galaxy}&\colhead{$N_H$}&\colhead{$A1$}
&\colhead{$E$\tablenotemark{a}$_{line}$}
&\colhead{$\sigma$\tablenotemark{b}$_{line}$}
&\colhead{$EW$\tablenotemark{c}$_{line}$}
&\colhead{$F_{2-10}$}&&\colhead{$N_{H}$}&\colhead{$A2$}&\colhead{$kT$}&\colhead{$A3$}
&\colhead{$F_{0.5-2}$}
&\colhead{$\chi^2/$dof}
}
\startdata
\sidehead{{\it ASCA}}
NGC 1068 & \nodata & \nodata &$6.4^{+0.03}_{-0.06}$ & 0.05f & $1000^{+260}_{-630
}$& $49\pm 1.7$ && 3.0f& $9.0^{+0.50}_{-0.56}$& $0.69\pm 0.02$& $18^{+0.39}_{-0.77}$& $64.2\pm 3.2$&  1870/786 \\
         &\nodata & \nodata &$6.6^{+0.07}_{-0.06}$&0.50f&$4200^{+500}_{-630}$&\nodata&&\nodata&\nodata&$1.66\pm0.2$&$7.6^{+1.6}_{-1.3}$&\nodata&\nodata\\
         &\nodata & \nodata &$6.7^{+0.24}_{-0.08}$&0.05f&$350^{+150}_{-190}$&\nodata&&\nodata&\nodata&\nodata&\nodata&\nodata&\nodata\\
Mrk 1066 & $57^{+59}_{-45:}$ & $0.97^{+0.25}_{-0.29}$ & $6.6^{+0.20}_{-0.52}$ & 0.05f & $3200\pm1900$ & $3.6\pm0.47$ && 12f & \nodata & $0.88^{+0.22}_{-0.20}$& $0.42^{+0.14}_{-0.11}$& $1.3\pm0.17$& 74/73 \\
Mrk 1073 & $400^{+390}_{-220}$&$2.2^{+0.89}_{-0.70}$&\nodata&\nodata&\nodata&$4.7^{+0.74}_{-0.85}$&&14f&\nodata&$1.0^{+0.47}_{-0.29}$&$0.70^{+0.82}_{-0.28}$&$1.3\pm0.26$&119/63\\
IC 3639 &$2100^{+3000}_{-1300}$&$3.4^{+ 3.2}_{-1.7}$&\nodata&\nodata&$<830$&$ 4.8\pm1.2$&&5.1f&\nodata&$ 2.3^{+2.9}_{-0.7}$&$1.7\pm0.53$&$1.2\pm0.29$& 48/53 \\
NGC 5135 &$ 12900^{+16000}_{-6800}$&$ 26^{+120}_{-18}$& \nodata&\nodata&$<1100$&$ 6.2^{+1.5}_{-1.6}$&& 4.7f &$ 0.67^{+0.09}_{-0.15}$&$ 0.77^{+0.04 }_{-0.1}$&$ 0.79^{+0.28}_{-0.14}$&$ 2.8^{+0.5}_{-0.2}$& 108/98 \\ 
Mrk 266 & $160^{+330}_{-150}$&$1.2^{+1.1}_{-0.6}$&\nodata&\nodata&$<10000$&$3.1\pm1.1$&&1.5f&\nodata&$0.78\pm0.3$&$0.73^{+0.35}_{-0.39}$&$1.8\pm0.6$&13/33\\
Mrk 273 & $3300^{+2900}_{-1100}$& $3.9^{+2.3}_{-1.9}$& $6.5^{+0.21}_{-0.17}$ & 0.05f&$860^{+130}_{-230}$&$5.2\pm1.0$&& 1.0f &$0.48^{+0.14}_{-0.21}$&$1.1^{+0.88}_{-0.77}$&$0.19^{+0.41}_{-0.17}$ &$1.2\pm0.24$& 55/57 \\ 
Mrk 463 &$ 2900^{+1700}_{-1200}$&$ 6.0^{+4.7}_{-2.8}$& \nodata&\nodata&$<890$&$ 6.6\pm1.3$&&2.1f&$ 0.40^{+0.080}_{-0.086}$&$ 0.71^{+0.53}_{-0.45}$&$ 0.15^{+0.080}_{-0.11}$&$ 1.1\pm0.17$& 80/70 \\
Mrk 477 & $2400^{+1700}_{-1200}$&$8.8^{+8.9}_{-4.3}$&$6.4^{+0.23}_{-0.21}$&0.05f&$560^{+560}_{-500}$&$12^{+2.3}_{-1.9}$&& 1.3f & $0.59^{+0.22}_{-0.20}$& \nodata&\nodata&$1.2\pm0.2$&36/104\\ 
NGC 6221 & $110^{+8.6}_{-8.3}$&$46^{+1.9}_{-1.3}$&$6.6^{+0.31}_{-0.29}$&0.50f&$360^{+210}_{-93}$&$140\pm4.1$&&15f&$3.0^{+1.4}_{-1.5}$&$1.4^{+1.6}_{-0.5}$&$1.2^{+1.7}_{-0.69}$&$24\pm0.7$&799/765\\
NGC 7130 &$ 16^{+62}_{-14:}$&$ 0.60\pm0.3$&\nodata&\nodata&$<5400$&$ 1.8\pm0.3$&&2.0f&\nodata&$ 0.79^{+0.2 }_{-0.1}$&$ 0.56^{+0.29}_{-0.25}$&$ 2.0^{+0.3}_{-0.4}$& 74/74 \\
NGC 7582 (1994) & $1100^{+74}_{-78}$& $84^{+6.6}_{-6.5}$& $6.2^{+0.17}_{-0.33}$ & 0.05f&$190^{+60}_{-140}$&$140\pm7.1$&& 1.5f &$2.0^{+0.33}_{-0.30}$&\nodata&\nodata&$4.4^{+0.6}_{-0.5}$& 344/284\\ 
NGC 7582 (1996)&$1400^{+60}_{-50}$& $93^{+4.6}_{-4.4}$& $6.3\pm0.08$ & 0.05f&$150^{+53}_{-51}$&$140\pm 4.2 $&& 1.5f &$2.1^{+0.14}_{-0.16}$&\nodata&\nodata&$4.4^{+0.2}_{-0.3}$& 516/467\\ 
\sidehead{{\it ASCA + PSPC}}
NGC 1068 & \nodata & \nodata &$6.4\pm 0.05$  & 0.05f & $1000\pm 310$& $50\pm 2.5$ && 3.0f& $11^{+0.27}_{-0.29}$& $0.79\pm 0.01$& $18^{+0.55}_{-0.46}$& $100\pm 5.0$&  2315/910 \\
         &\nodata & \nodata &$6.6\pm0.07$&0.50f&$3500^{+250}_{-390}$&\nodata&&\nodata&\nodata&$0.13\pm0.005$&$110^{+9.2}_{-7.9}$&\nodata&\nodata\\
         &\nodata & \nodata &$6.7^{+0.10}_{-0.14}$&0.05f&$530^{+410}_{-280}$&\nodata&&\nodata&\nodata&\nodata&\nodata&\nodata&\nodata\\
NGC 5135 &$ 14000^{+16000}_{-3700}$&$ 29^{+100}_{-21}$&\nodata&\nodata&$<590$&
$ 6.3\pm0.63$&&4.7f&$ 0.71^{+0.11}_{-0.10}$&$ 0.68^{+0.14}_{-0.06}$&$ 0.79^{+0.10}_{-0.11}$&$ 2.9\pm0.29$& 150/119 \\
Mrk 266 &$1.5:^{+1.1}_{-0:}$&$0.75\pm0.2$ &\nodata&\nodata&$<9400$&$2.1^{+0.4}_{-0.8}$&&1.5f&\nodata&$0.28^{+0.1}_{-0.05}$& $0.72^{+0.42}_{-0.33}$&$2.6^{+0.5}_{-0.9}$&43/50\\
Mrk 273 & $2500^{+1500}_{-890}$&$3.2^{+2.9}_{-1.4}$&$6.5^{+0.21}_{-0.17}$&0.05f&$970^{+630}_{-680}$&$4.8\pm0.91$&&1.0f&$0.35^{+0.082}_{-0.097}$&$0.84^{+0.48}_{-0.24}$&$0.28^{+.17}_{-0.097}$&$1.3^{+0.15}_{-0.19}$&90/68\\
Mrk 463 &$ 2800^{+1300}_{-850}$&$ 5.8^{+5.0}_{-2.7}$&\nodata&\nodata&$<960$&$ 6.6\pm1.3$&&2.1f&$ 0.40^{+0.085}_{-0.094}$&$ 0.66^{+0.35}_{-0.39}$&$ 0.16^{+0.086}_{-0.084}$&$ 1.1\pm0.17$& 96/87  \\
NGC 7582 (1994) &  $1000\pm70$& $84^{+6.4}_{-5.9}$& $6.2^{+0.18}_{-0.36}$ & 0.05f &$160^{+100}_{-84}$&$140^{+6.9}_{-6.1}$&&$5.3^{+5.7}_{-1.6}$&$1.5^{+0.44}_{-0.32}$&$0.46^{+0.32}_{-0.19}$ & $0.56^{+0.61}_{-0.32}$&$4.10^{+0.9}_{-1.0}$&365/297\\ 
NGC 7582 (1996)& $1400\pm50$& $93^{+4.6}_{-4.2}$& $6.3\pm0.08$ & 0.05f &$140^{+55}_{-48}$&$140^{+4.3}_{-4.0} $&&$5.6^{+3.4}_{-1.4} $&$1.9\pm0.20$&$0.60^{+0.24}_{-0.30}$&$0.32^{+0.16}_{-0.17}$ &$4.2\pm0.4$&538/480 \\ 

\enddata

\tablenotetext{a}{Energy of line center in keV.}
\tablenotetext{b}{Line width in keV.}
\tablenotetext{c}{Equivalent width of line in eV.}
\tablecomments{Power law photon index is fixed at 1.9.
Additional fixed parameters are marked with f.   
Other symbols and errors are described in 
Notes to Tables \ref{tab:pl}--\ref{tab:plm}.}

\end{deluxetable}

\begin{deluxetable}{llllc}
\tablewidth{0pt}
\tablecaption{Spectral Fits to PSPC Data\label{tab:pspc}}
\tablehead{
\colhead{Galaxy}&\colhead{$kT$ or $\Gamma$} 
&\colhead{$A$\tablenotemark{a}} 
&\colhead{$F_{0.5-2}$}&\colhead{$\chi^2/$dof}
}

\startdata
Mrk 78 & $kT=0.76\pm0.3$ & $3.2\pm1.0$ & $0.060^{+0.1}_{-0.2}$ & 16/7\\
NGC 7674 & $\Gamma = 2.0^{+0.6}_{-2.0}$ & $1.3^{+6.8}_{-5.4}$ & $2.3^{+1.0}_{-0.7}$ & 3/2 \\
\enddata

\tablenotetext{a}{Normalization in units of $10^{-4}$.}
\tablecomments{
Column density is fixed at Galactic value.
Other symbols and errors are described in 
Notes to Tables \ref{tab:pl} and \ref{tab:plm}.
}

\end{deluxetable}

\includegraphics[height=8in]{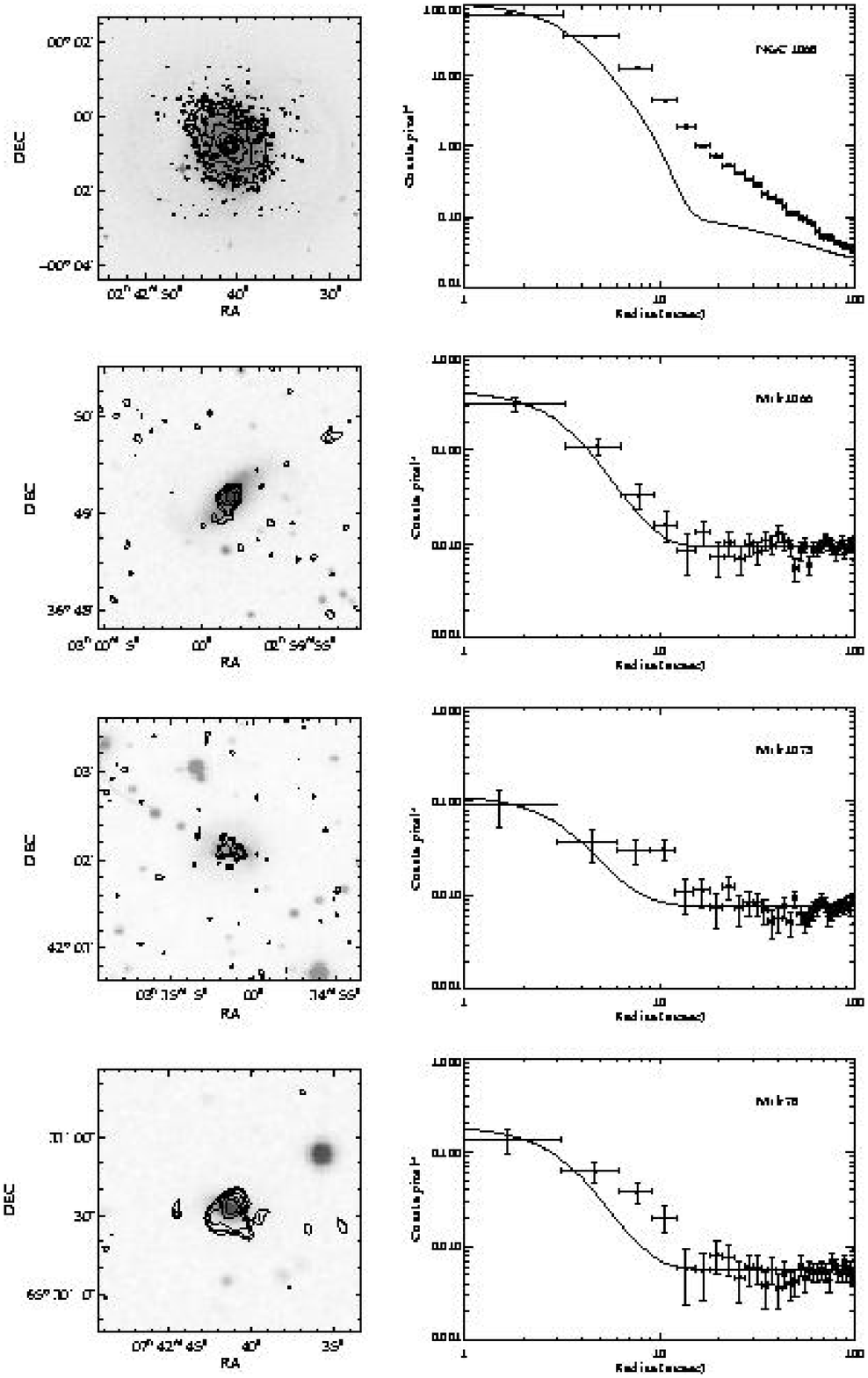}
\clearpage
\includegraphics[height=8in]{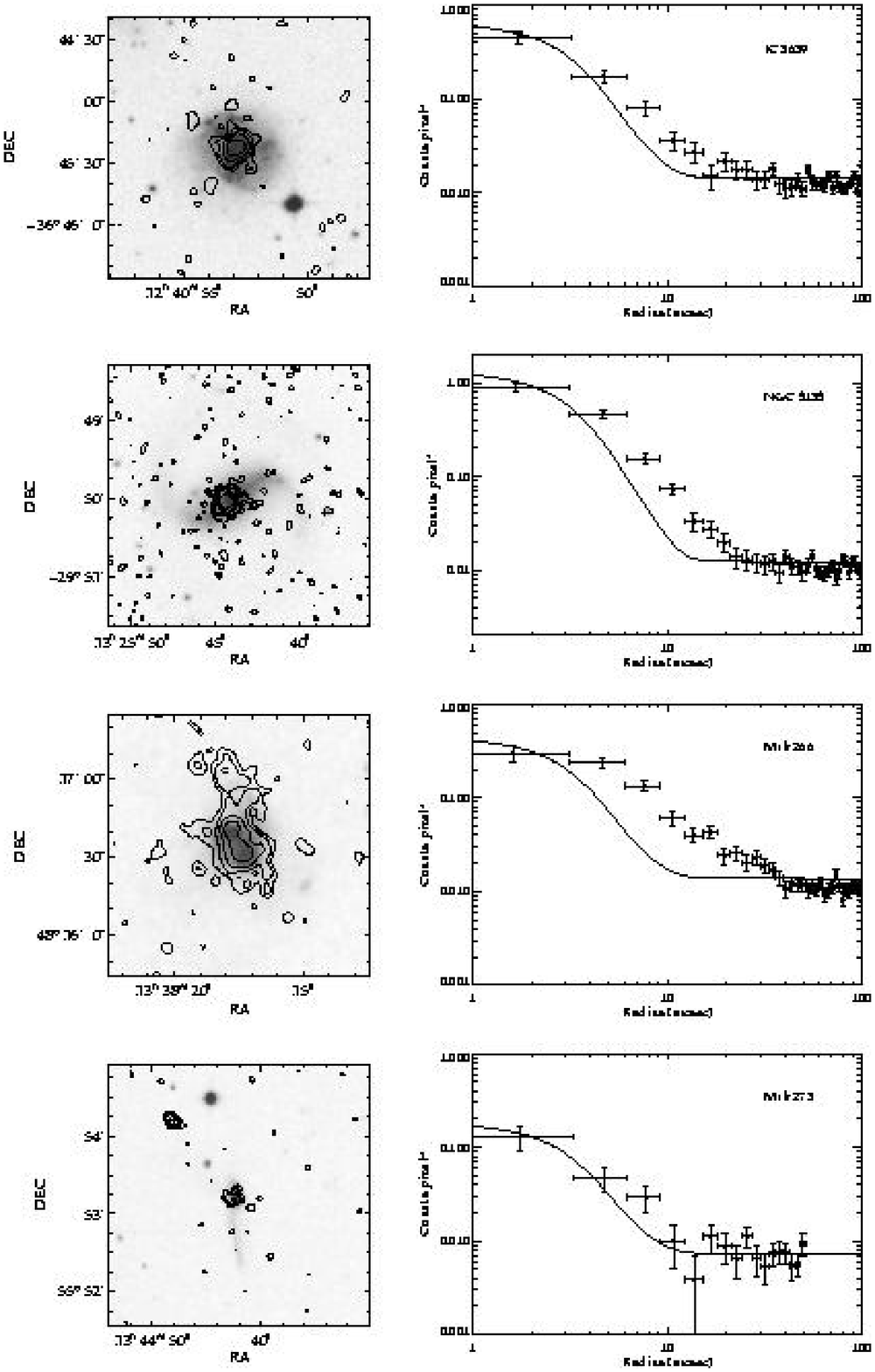}
\clearpage
\includegraphics[height=8in]{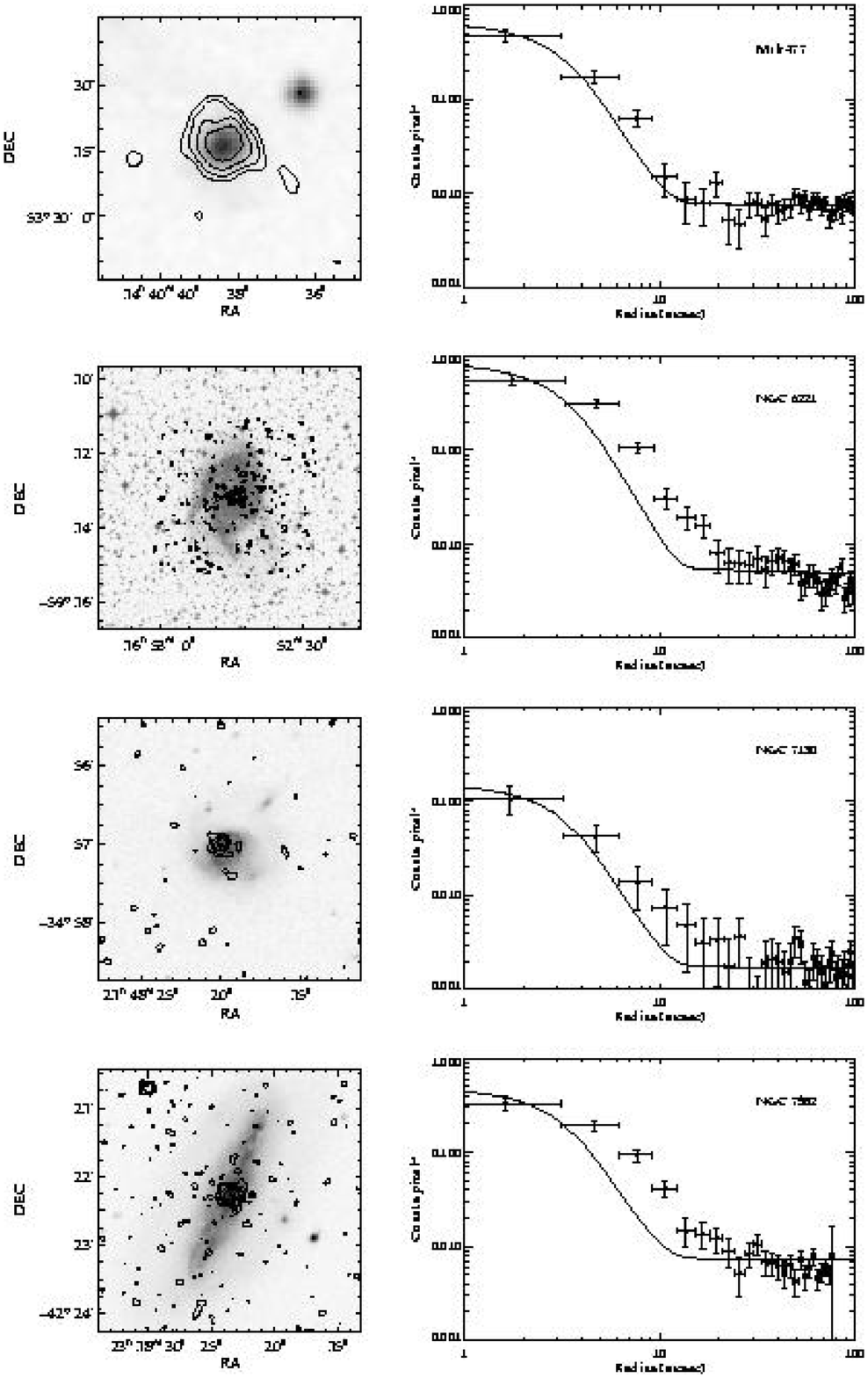}
\figcaption{Contours of {\it ROSAT} HRI intensity overlaid on DSS
images and radial profiles of HRI soft X-ray emission for the Sy2/SB sample.
Raw X-ray data have been smoothed by
a Gaussian of FWHM=4\arcsec.
The minimum contour level is 
3 standard deviations 
above the background, and 
contour intervals are logarithmic, in factors of 2, except where noted.
Radial profile data points are azimuthal averages 
of counts per raw ($0\farcs5$) pixel.  The background
and amplitude are fit to the theoretical
point-spread function ({\it solid line}).
NGC 1068, with scale factor 3 in contour levels; Mrk 1066; 
Mrk 1073, scale factor 1.5; 
Mrk 78, scale factor 1.5; IC 3639; NGC 5135; Mrk 266; Mrk 273;
Mrk 477, with minimum contour level 4$\sigma$; NGC 6221; NGC 7130, 
minimum contour level 5$\sigma$; NGC 7582,  minimum contour 
level 3.5$\sigma$.\label{fig:hriall}}
\clearpage

\includegraphics[height=4in]{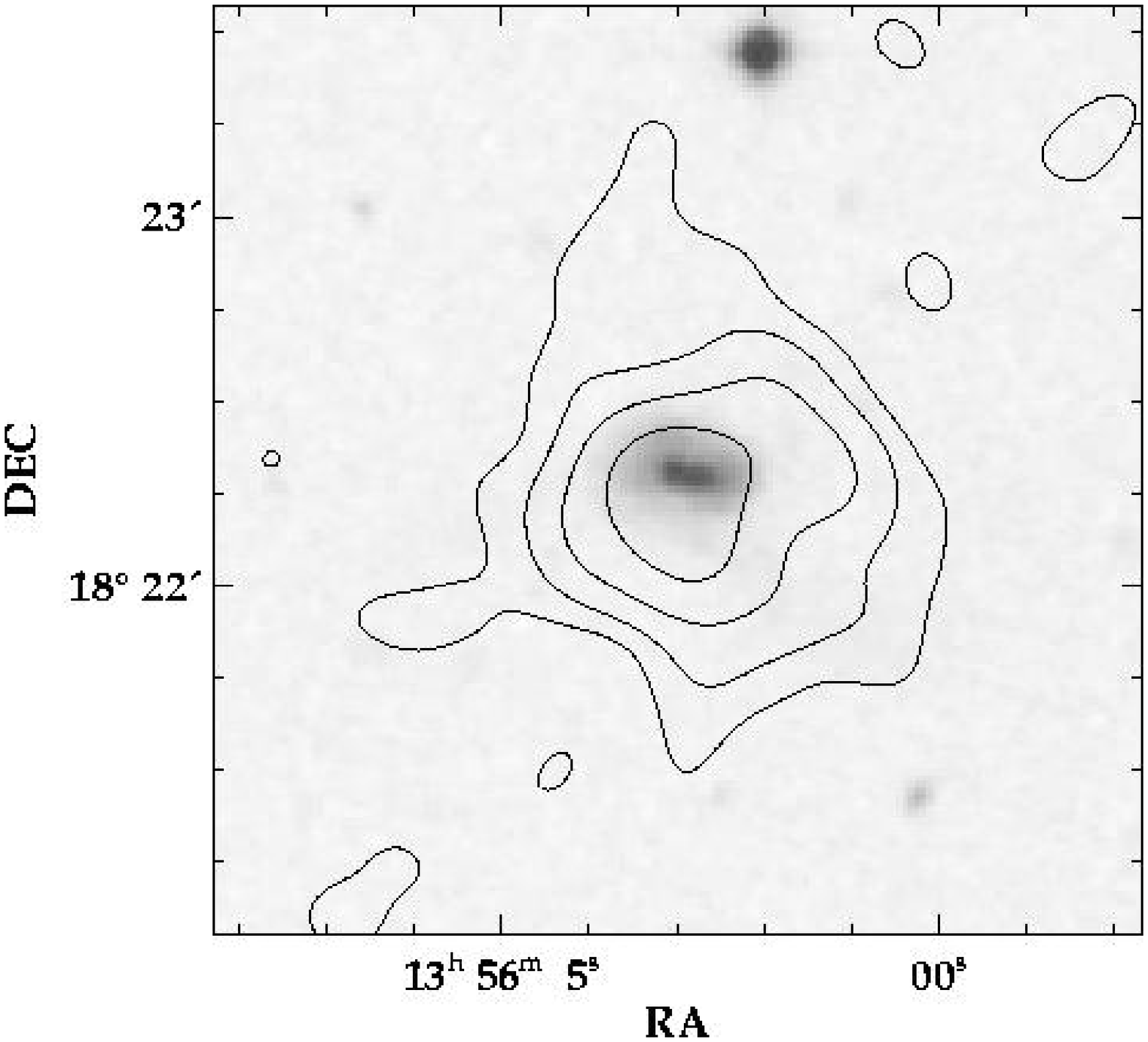}
\figcaption
{Contours of PSPC intensity overlaid on
DSS image of Mrk 463.  The X-ray data have been smoothed
by FWHM=15\arcsec.  The contour intervals are logarithmically spaced
by factors of 2, beginning with 3$\sigma$ above the background.
\label{fig:poptm463}}
\vskip 0.1in

\includegraphics[height=4in]{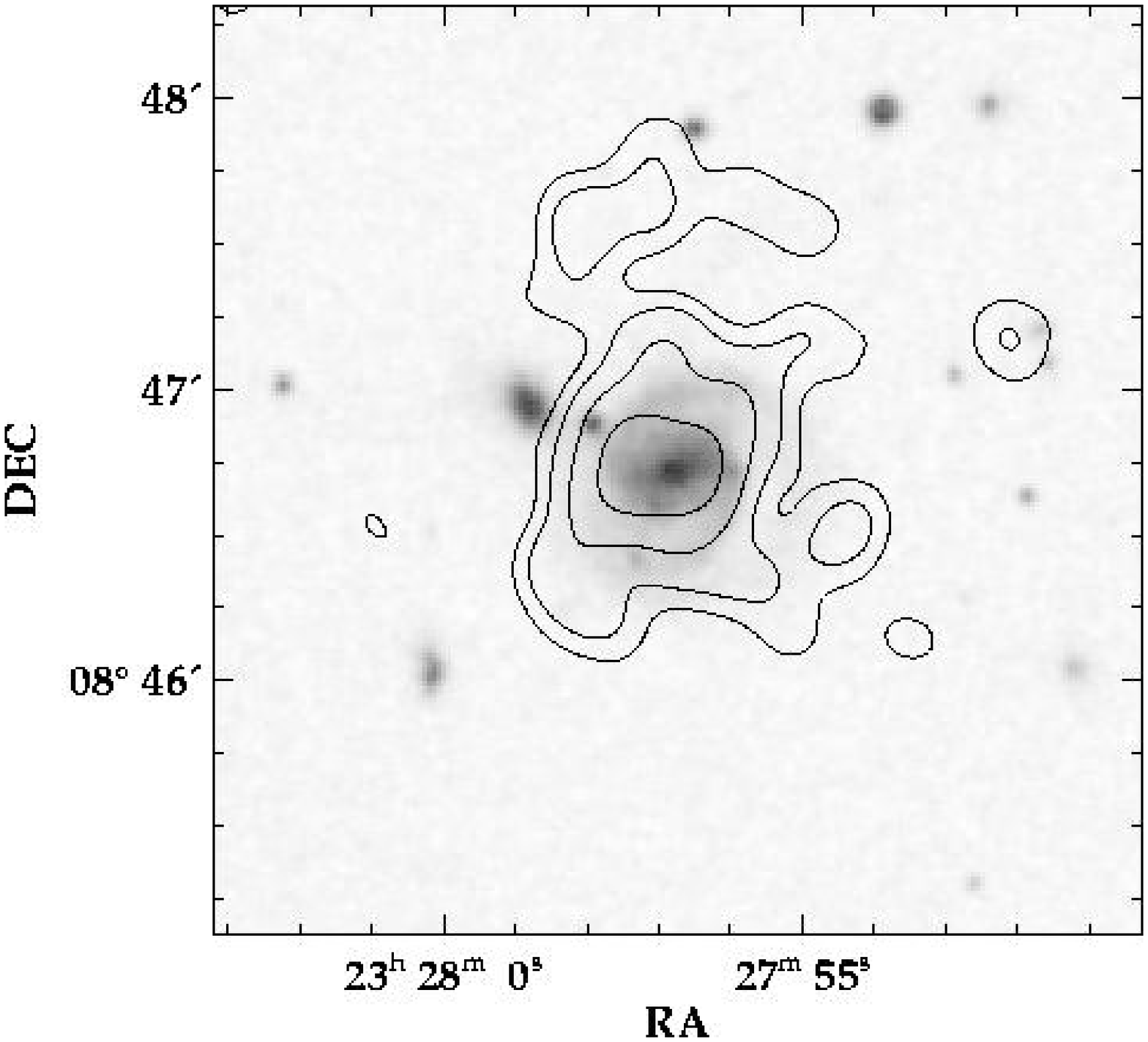}
\figcaption
{PSPC contours overlaid on DSS image of NGC 7674.
The X-ray data have been smoothed by FWHM=15\arcsec. 
The lowest contour level is 3$\sigma$ above the background, and 
the interval is spaced logarithmically by factors of 2.
\label{fig:poptn7674}}
\clearpage

\includegraphics[height=8in]{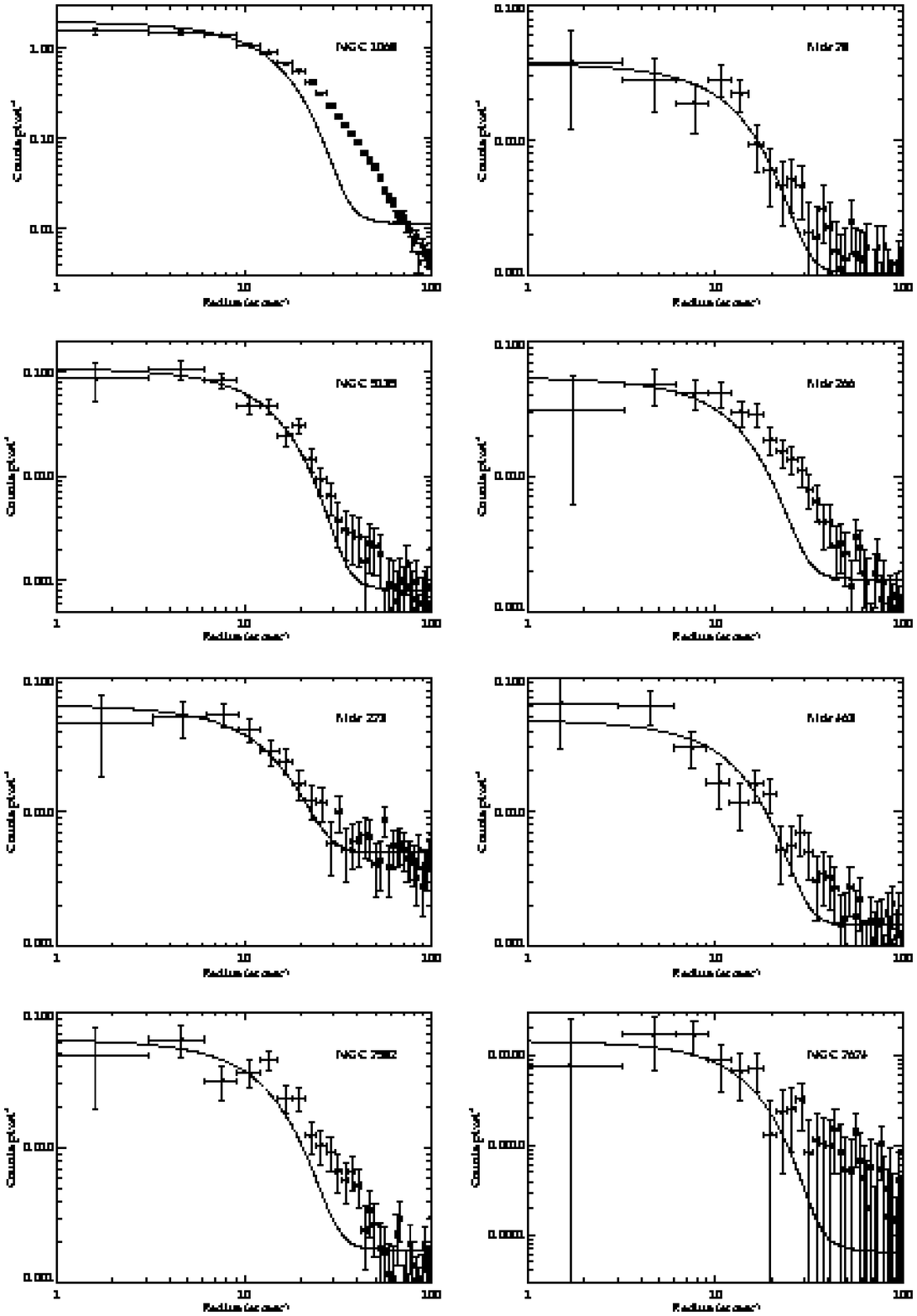}
\figcaption
{Radial profile of X-ray emission detected with {\it ROSAT} PSPC.
Data points are azimuthal averages 
of counts per raw ($0\farcs5$) pixel.  The background
and amplitude are fit to the theoretical
point-spread function ({\it solid line}).
As labelled, panels show NGC 1068, Mrk 78, NGC 5135, Mrk 266, Mrk 273, Mrk 463, 
NGC 7582, and NGC 7674 sequentially.
\label{fig:pspcall}
}
\clearpage

\includegraphics[width=6in]{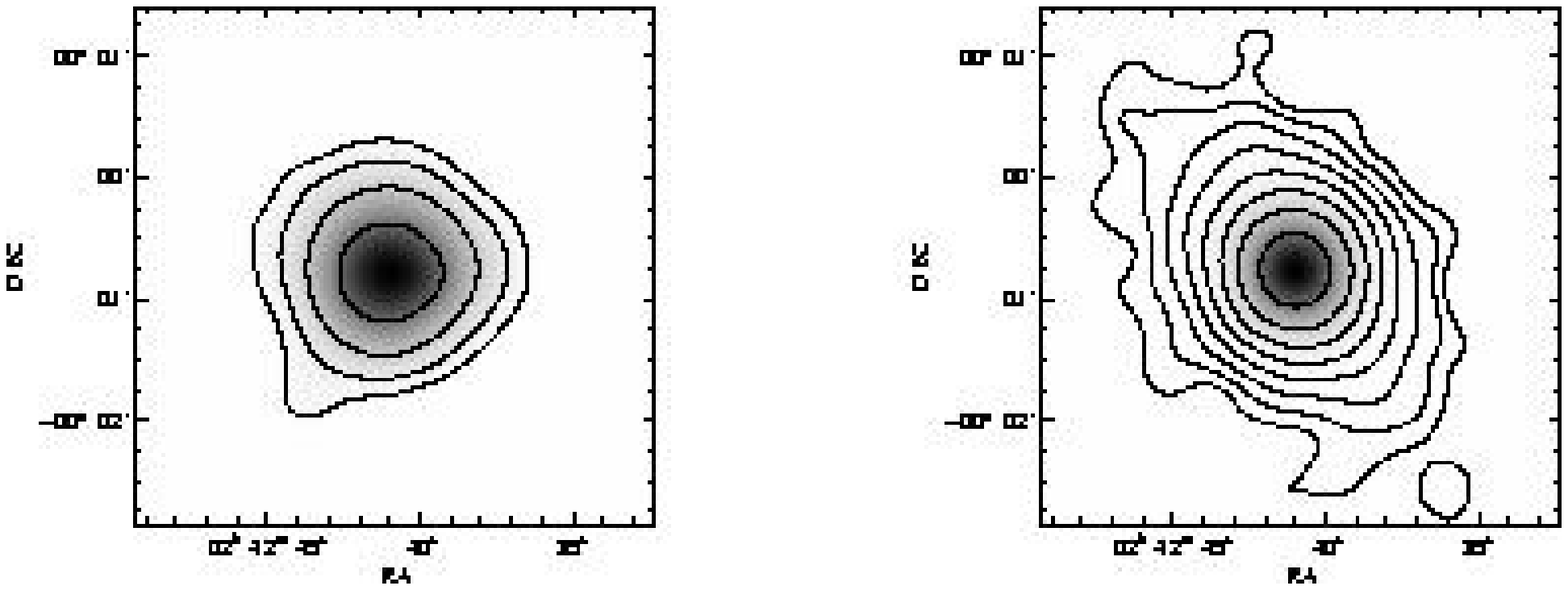}
\figcaption
{(a) Soft PSPC (0.1--0.4 keV) observation of NGC 1068, smoothed by
a Gaussian of FWHM=25\arcsec.  The image is scaled linearly from 
a minimum of 1$\sigma$ above the background. 
The lowest contour level is 2$\sigma$ above the background, and 
the interval is spaced logarithmically by factors of 2.5.
(b) Hard PSPC (0.4--2.1 keV) observation of NGC 1068, 
with the same image and contour scaling described in (a).
The spatial distributions of the 
two energy bands are distinct.  The harder emission is more centrally concentrated,
and the elongation directions are nearly
perpendicular to one another.
\label{fig:phsn1068}
}
\vskip 0.1in

\includegraphics[width=6in]{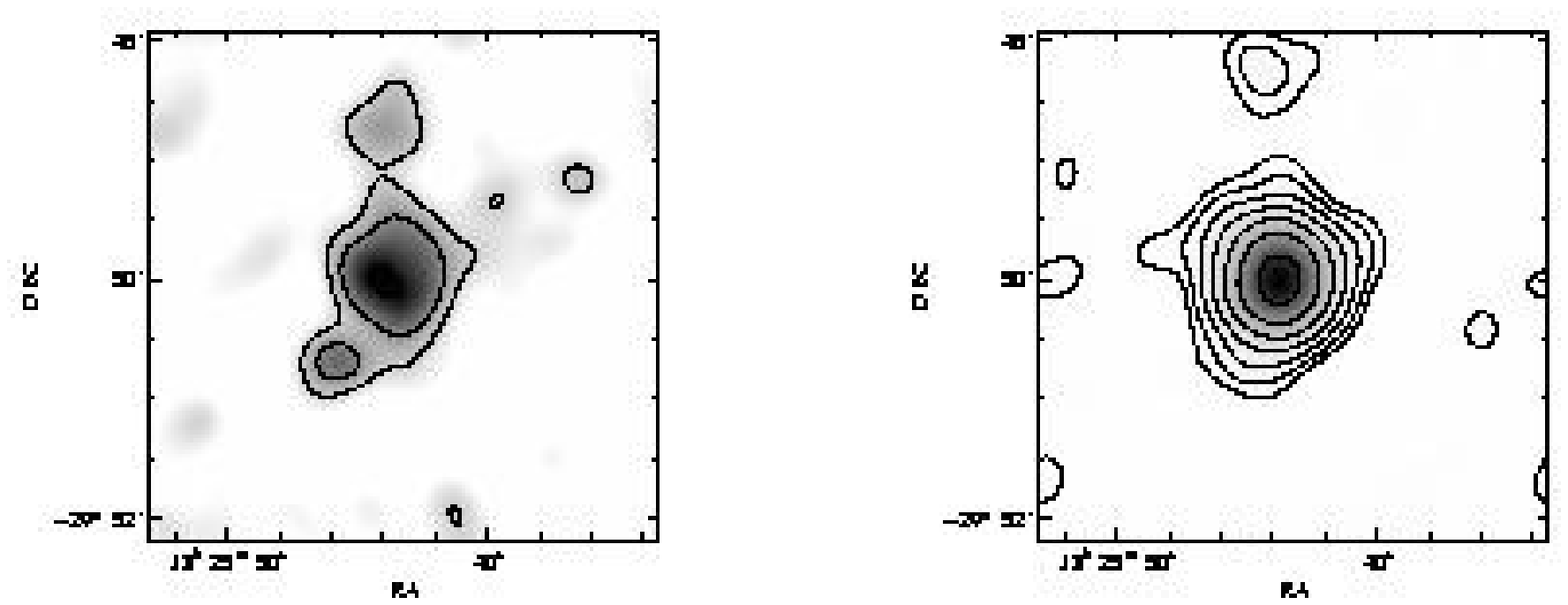}
\figcaption
{(a) Soft PSPC (0.1--0.4 keV) image of NGC 5135, smoothed by
a Gaussian of FWHM=25\arcsec\ and scaled linearly from 
a minimum of 1$\sigma$ above the background. 
The lowest contour level is 4$\sigma$ above the background, and 
the interval is spaced logarithmically by factors of 2.
(b) Hard PSPC (0.4--2.1 keV) image of NGC 5135, with smoothing, scaling
and contour levels as described in (a).
\label{fig:phsn5135}}

\vskip 0.1in
\includegraphics[width=6in]{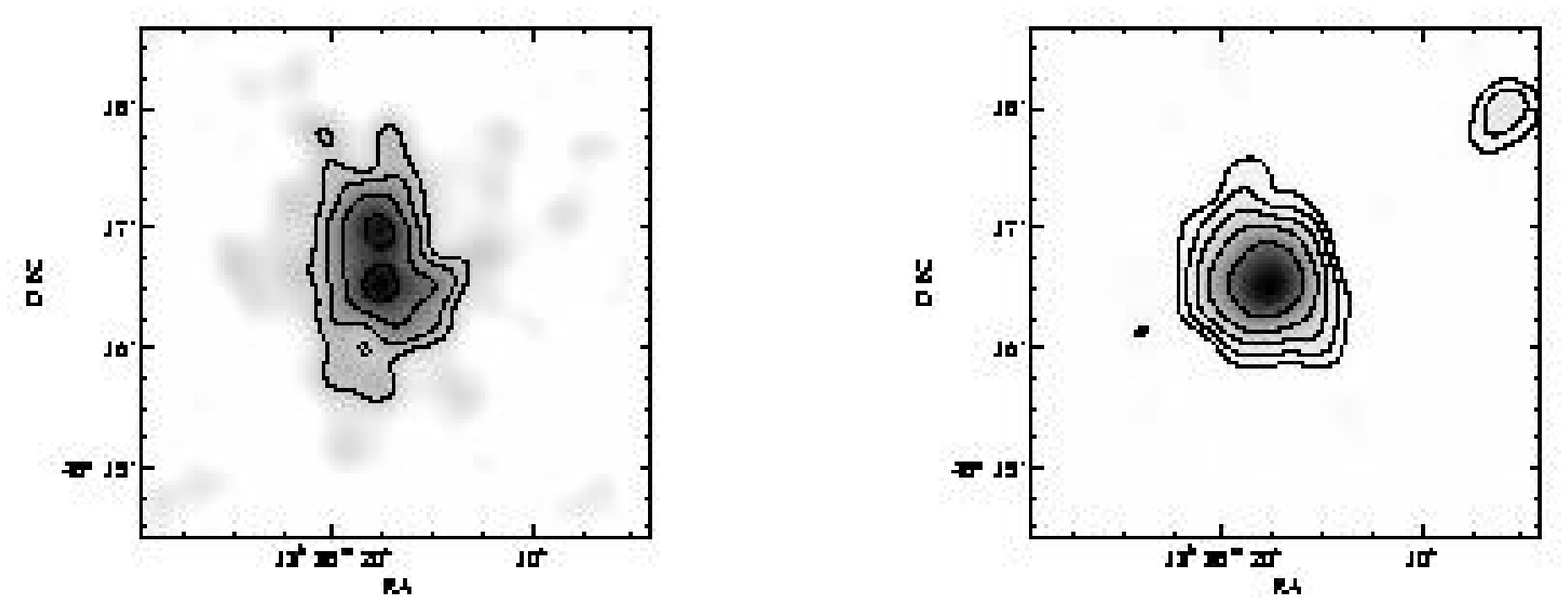}
\figcaption
{(a) Soft PSPC (0.1--0.4 keV) image of Mrk 266, smoothed by
a Gaussian of FWHM=20\arcsec.  The image is scaled linearly from 
a minimum of 1$\sigma$ above the background. 
The lowest contour level is 5$\sigma$ above the background, and 
the interval is spaced logarithmically by factors of 2.
(b) Hard PSPC (0.4--2.1 keV) observation of Mrk 266, 
with the same image and contour scaling described in (a).
The nuclei of Mrk 266 are the principal components of the harder emission, while
both the nuclei and the extended emission north of them are significant
at softer energies.
\label{fig:phsm266}
}
\clearpage

\includegraphics[width=6in]{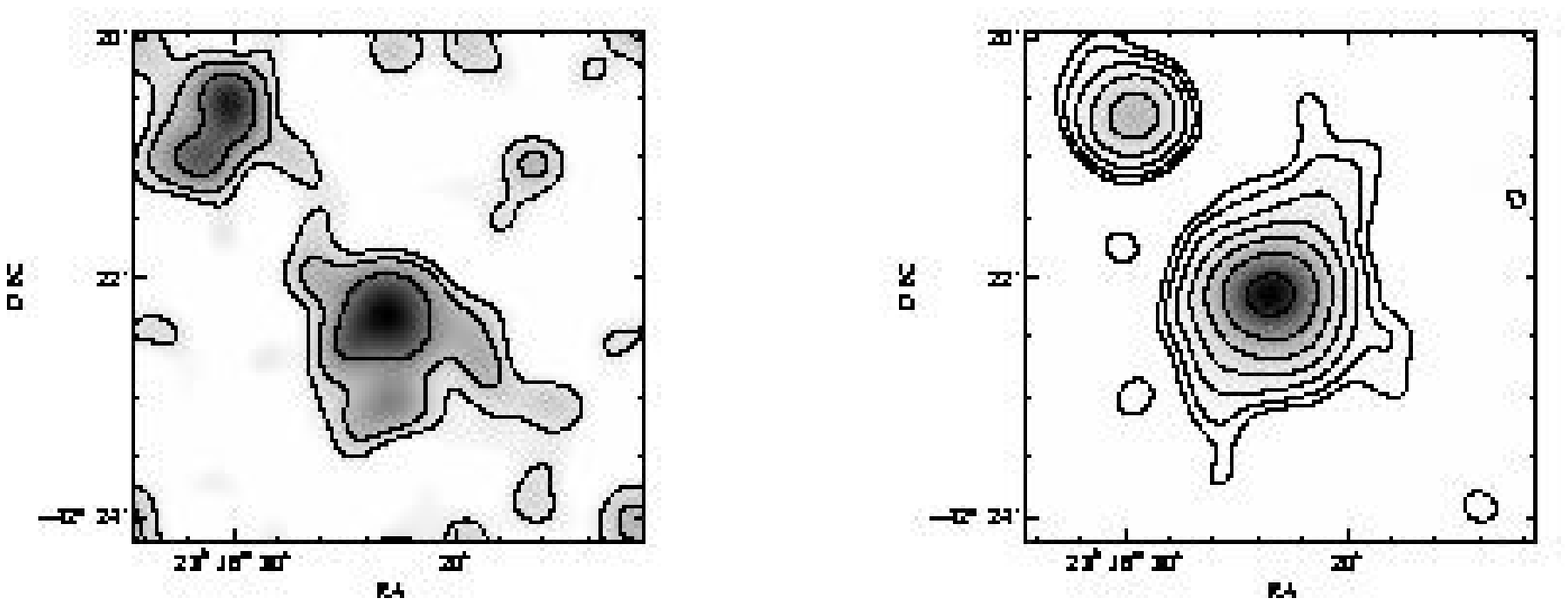}
\figcaption
{(a) Soft PSPC (0.1--0.4 keV) image of NGC 7582, smoothed by
a Gaussian of FWHM=25\arcsec.  The image is scaled linearly from 
a minimum of 1$\sigma$ above the background. 
The lowest contour level is 3$\sigma$ above the background, and 
the interval is spaced logarithmically by factors of 2.
(b) Hard PSPC (0.4--2.1 keV) observation of NGC 7582, 
with the same image and contour scaling described in (a).
The softest emission is extended preferentially perpendicular to 
the galactic disk.  The harder emission is more centrally concentrated
and more extended  in the plane of the galaxy.
\label{fig:phsn7582}
}

\vskip 0.1in
\includegraphics[width=6in]{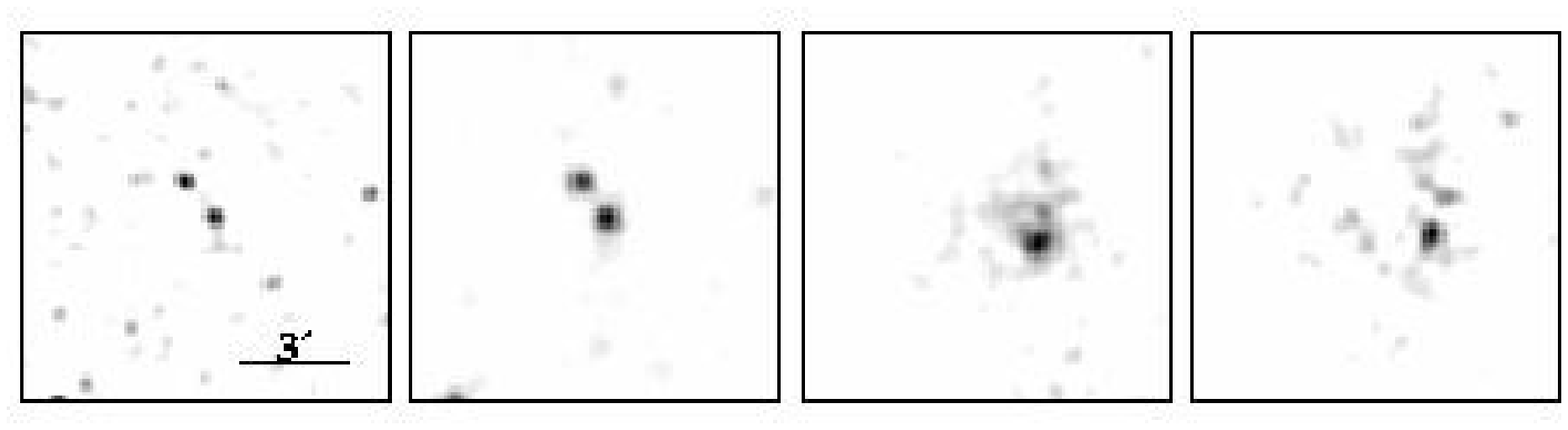}
\figcaption
{Mrk 273 in \rosat{} HRI, \rosat{} PSPC (0.5--2 keV), \asca{} SIS0 soft 
(0.5--2.0 keV), and \asca{} SIS0 hard (2.0-10 keV) bands.  
These approximately aligned images illustrate that the companion galaxy, located 
about 2\arcmin\ northeast of Mrk 273, faded 
from the time of the PSPC observation (1992 June) to the
time of the \asca{} observation (1994 May).  Although the extracted
\asca{} spectra include the companion, it does not contribute
significantly to the measured flux.
\label{fig:m273x4}
}

\vskip 0.1in
\includegraphics[width=6in]{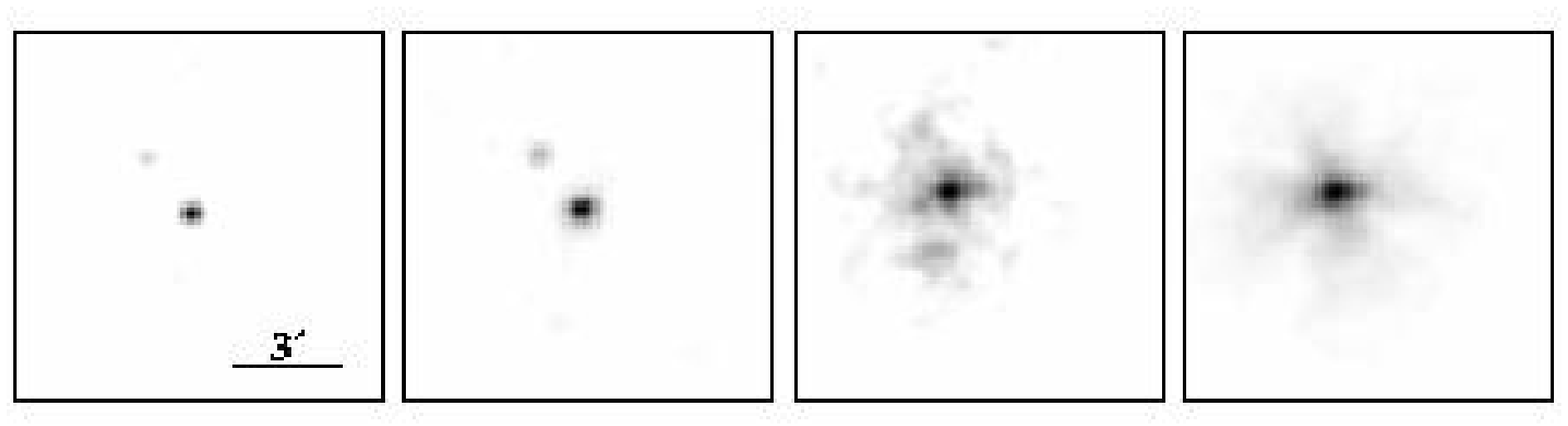}
\figcaption
{NGC 7582 in \rosat{} HRI, \rosat{} PSPC, \asca{} SIS0 soft, and
\asca{} SIS0 hard bands.  
The soft spectrum of the companion of NGC 7582, located about 2\arcmin\ 
toward the northeast, does not significantly contaminate the \asca{} 
spectra in the observation from 1994 or 1996 (shown).
\label{fig:n7582x4}
}
\clearpage

\includegraphics[height=8in]{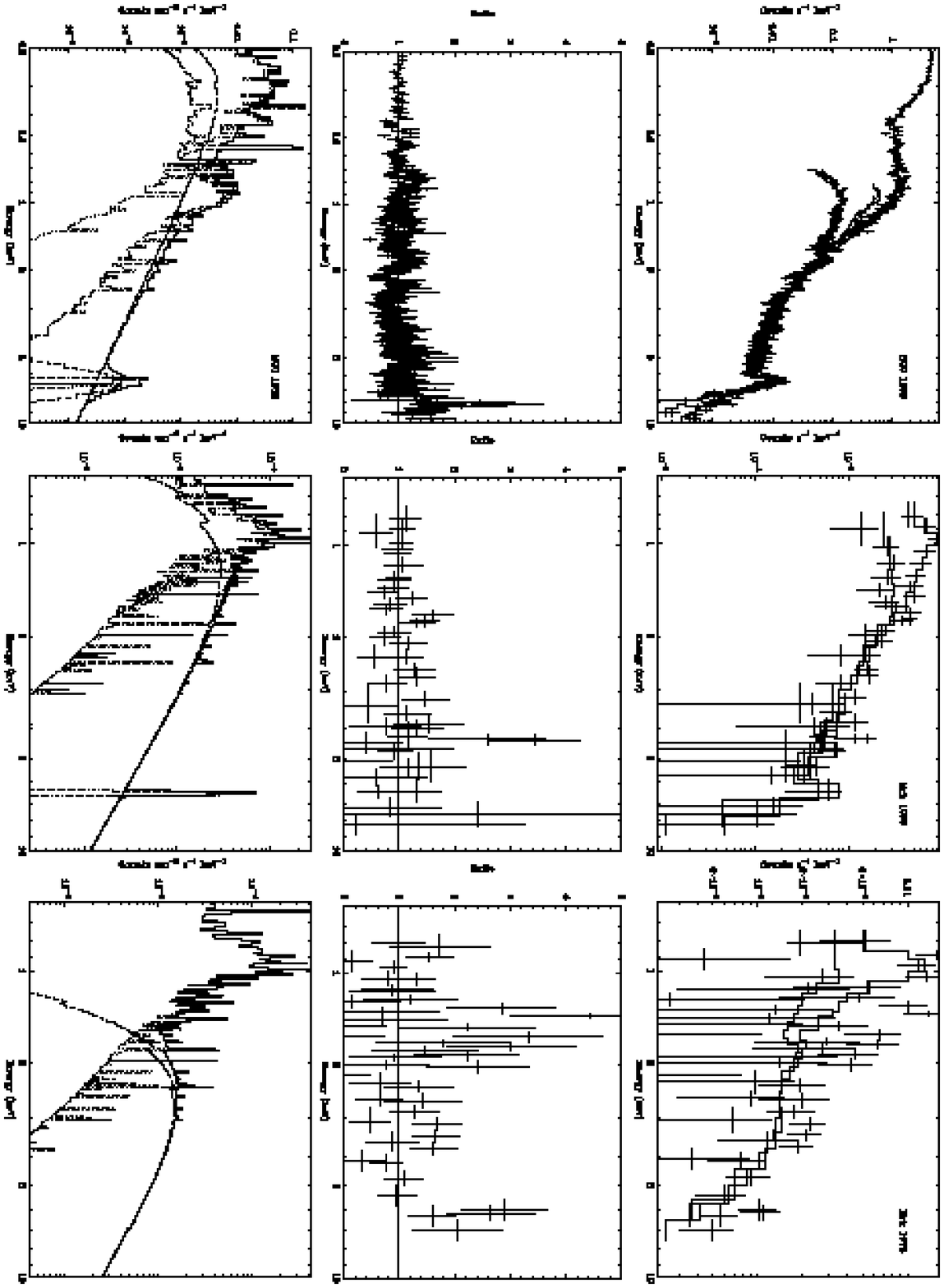}
\clearpage
\includegraphics[height=8in]{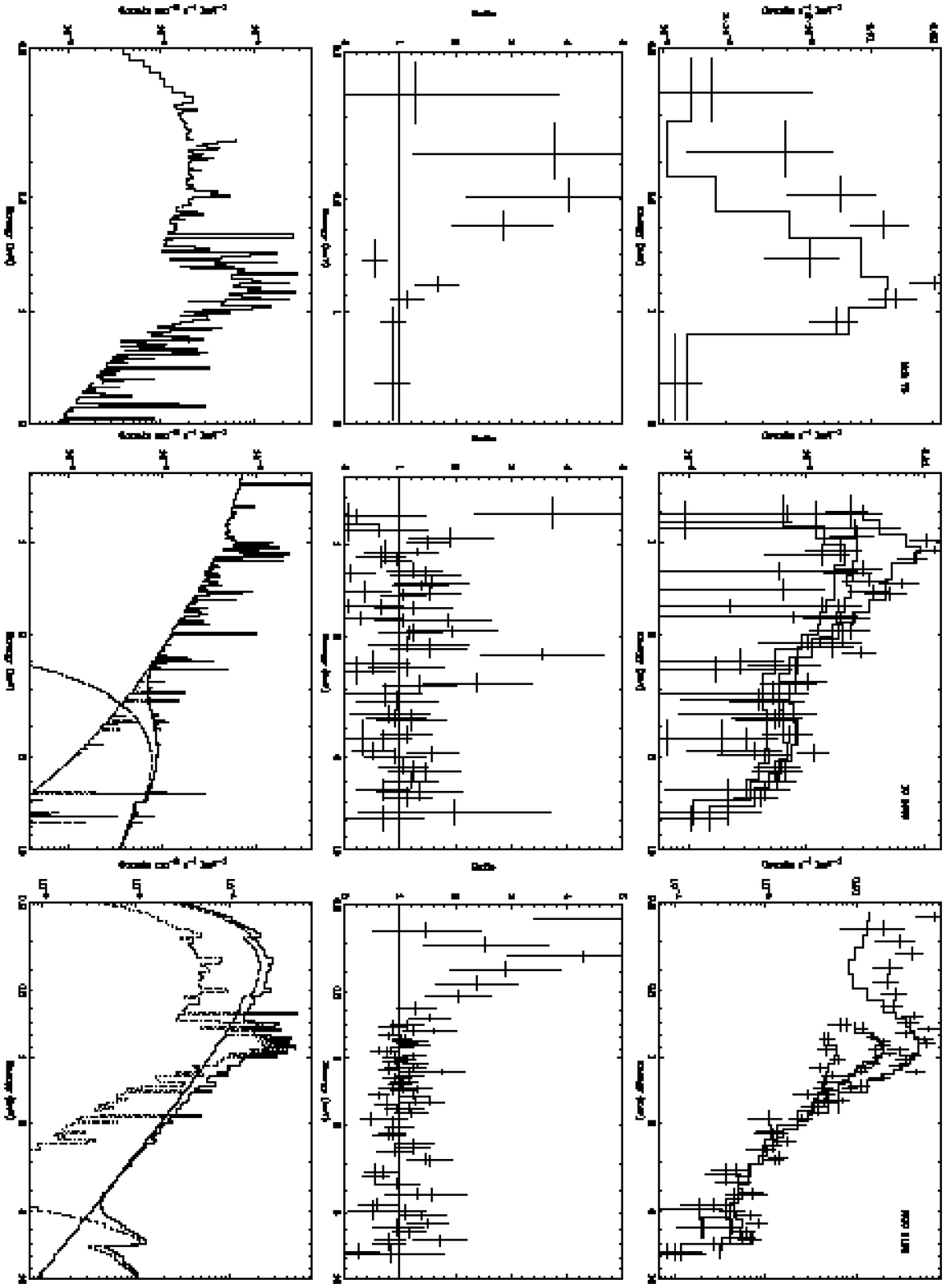}
\clearpage
\includegraphics[height=8in]{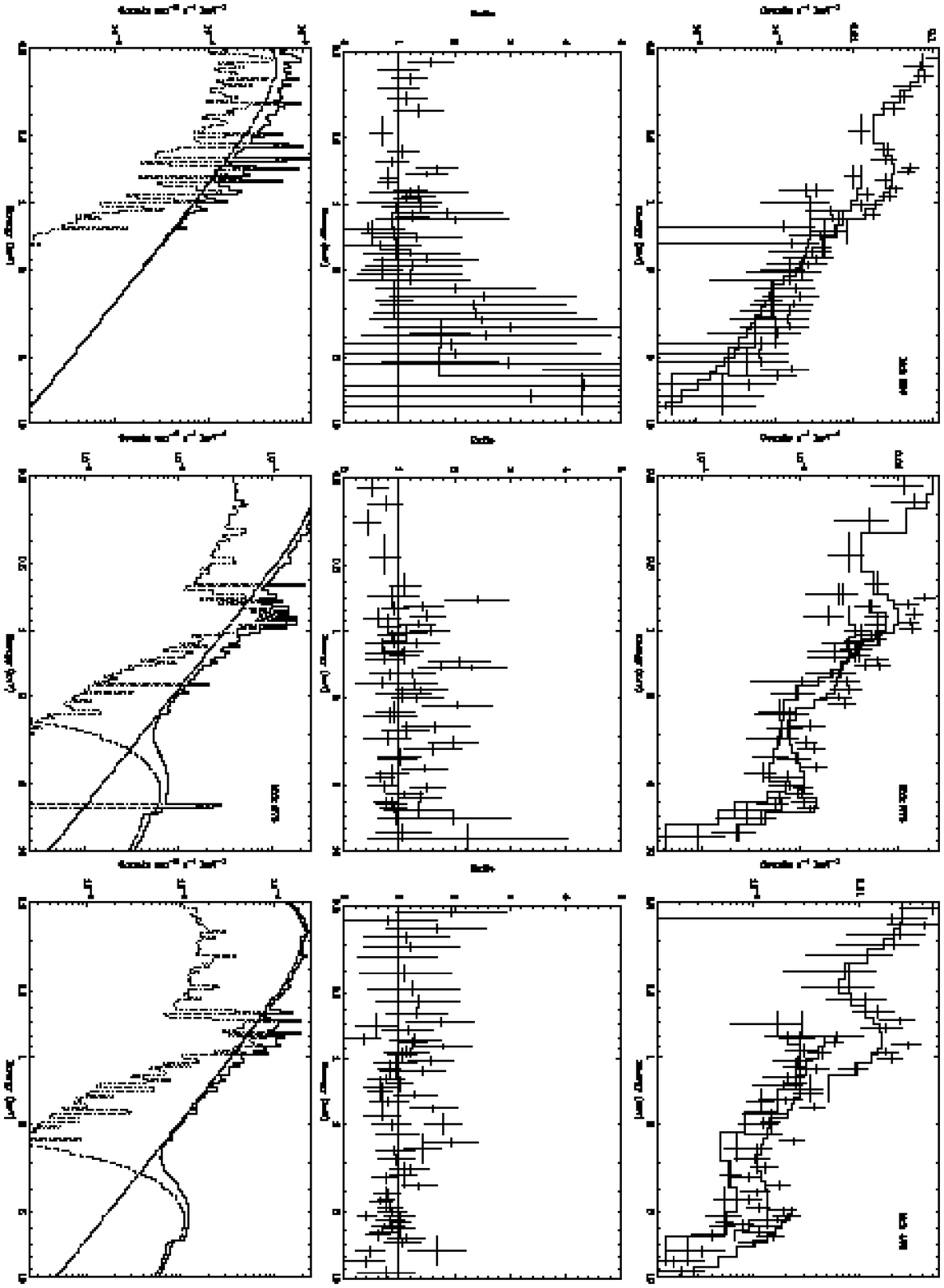}
\clearpage
\includegraphics[height=8in]{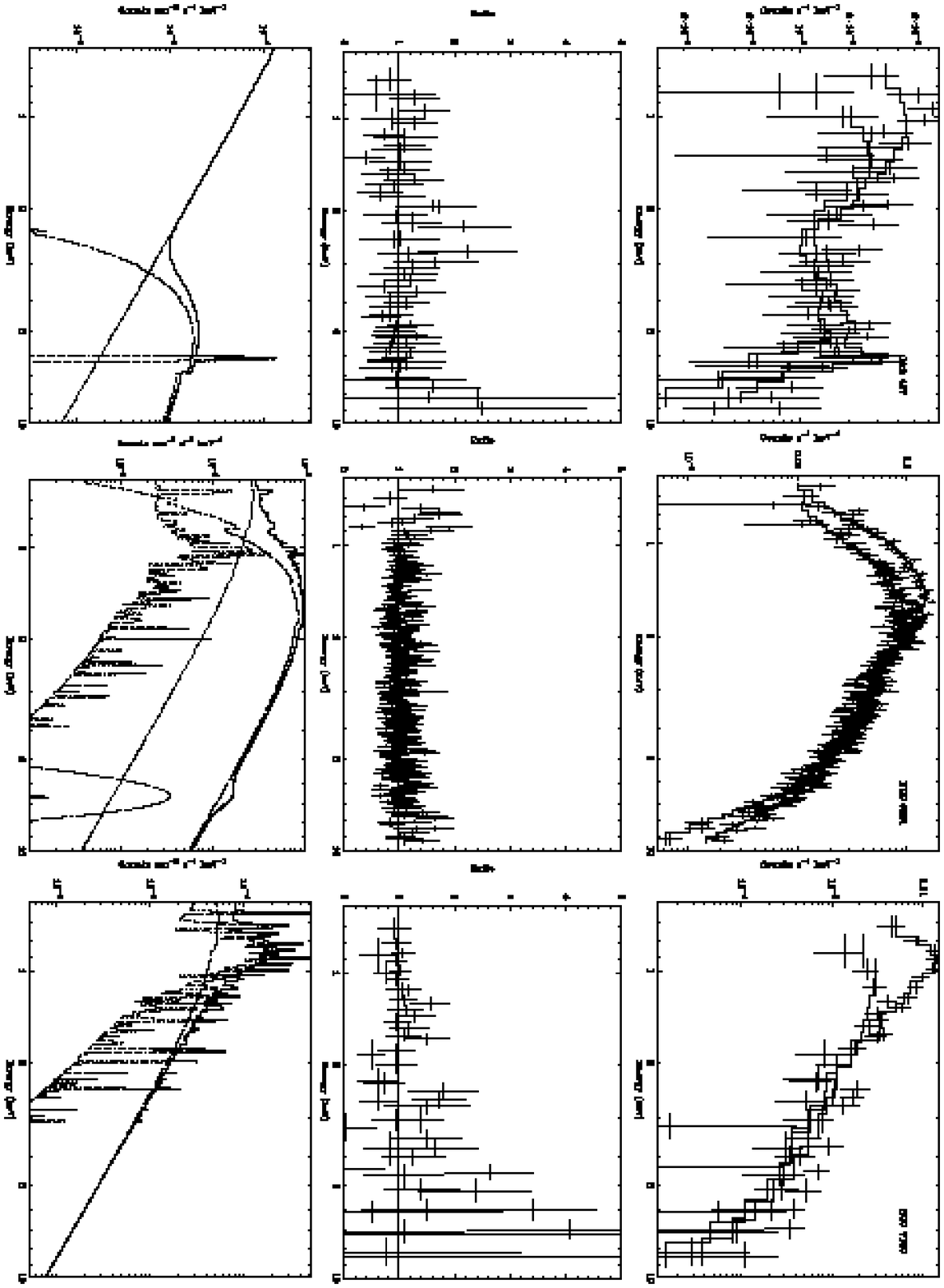}
\clearpage
\includegraphics[height=8in]{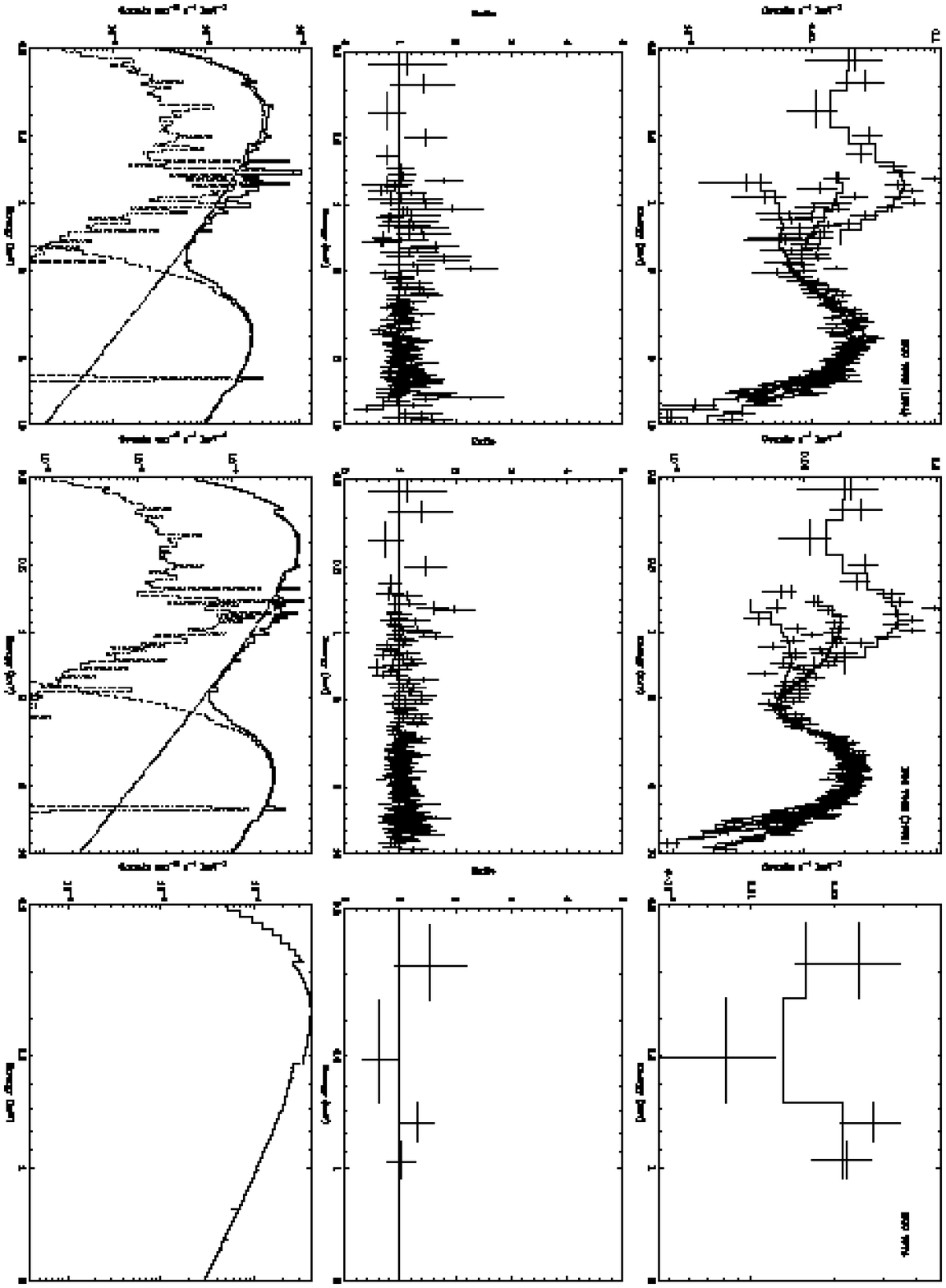}
\figcaption
{X-ray spectra of Sy2/SB composite galaxies.
For each galaxy, the top panel contains the data and best-fitting model 
folded through the instrument response.
The SIS0 and SIS1 data are
combined for plotting purposes, as are the GIS2 and GIS3 data.
PSPC data are also included in the relevant instances.
The second panel contains the the ratio of the data to the model.
The bottom panel contains the total model spectrum at higher resolution
({\it solid line}) and the individual model components ({\it broken lines}).
\label{fig:spectra}
}
\clearpage

\includegraphics[width=6in]{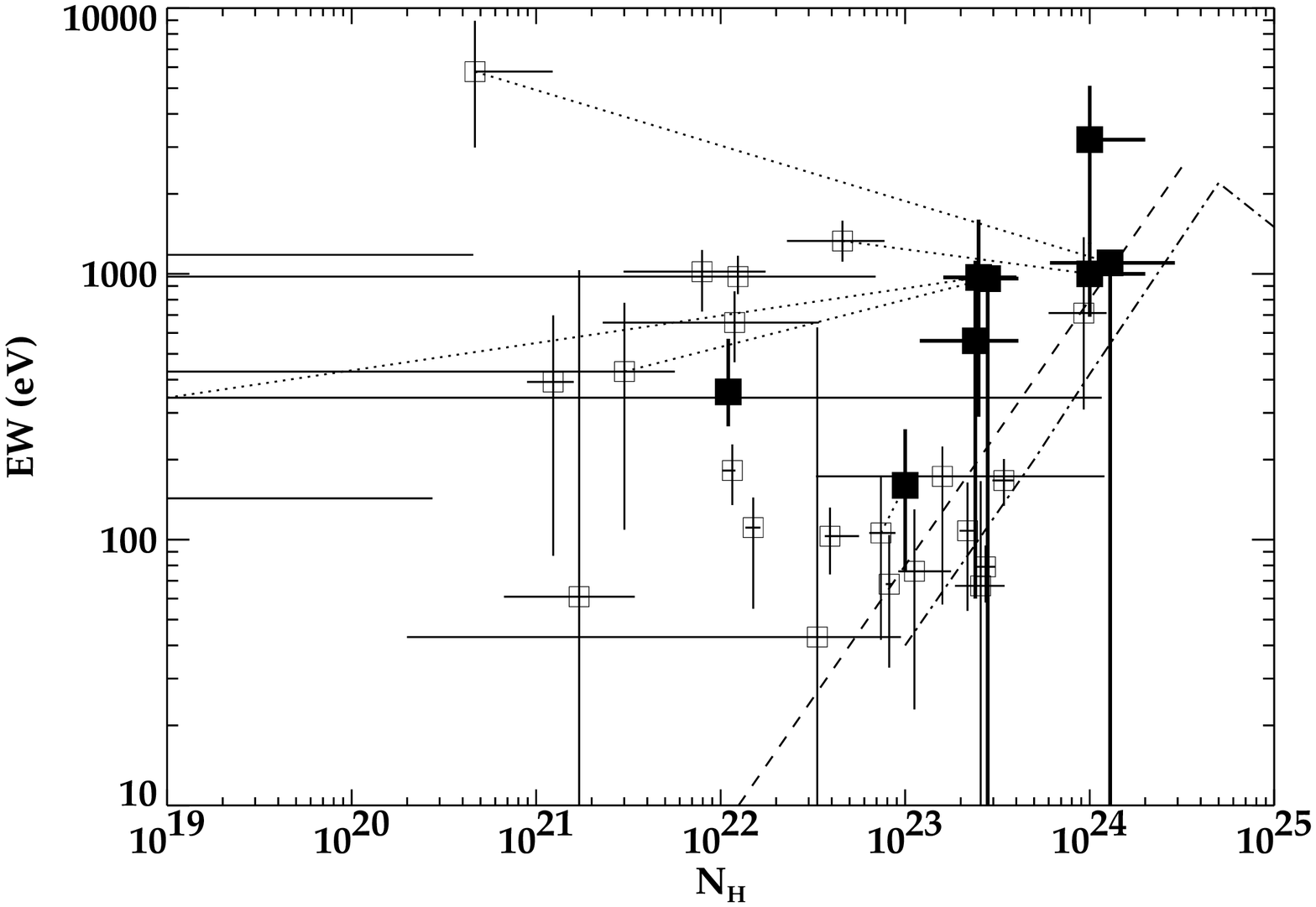}
\figcaption
{Fe K equivalent width vs. $N_H$ for our sample ({\it solid symbols}).
For comparison, we also plot the results from
Turner et al. (1997; {\it open symbols}),  %\nocite{Tur97}
which were derived using the observed 
continuum. Common objects in the two samples are connected with dotted
lines. Theoretical predictions for spherically-symmetric 
({\it dashed line}; \citealt{Awa91}) and torus 
({\it dot-dashed line}; \citealt{Ghi94}) 
geometries are also plotted.  Although we cannot distinguish between
these two models, our results are generally consistent with both of them.
\label{fig:ewnh}
}

\end{document}